%% file: as2.tex
\documentclass[12pt]{article}
\usepackage{fullpage}
\usepackage{setspace}
\usepackage{epsf}
\usepackage{natbib}
\usepackage{epsfig}
\usepackage{amsthm}
\usepackage{color}
\usepackage{amsmath}
\usepackage{amssymb}

\newcommand{\bm}[1]{\mbox{\boldmath $#1$}}
\newcommand{\mb}[1]{\mathbf{#1}}

\begin{document}

\doublespacing

\title{
Adaptive design and analysis of supercomputer experiments
}

\author{
\vspace{-0.2cm}
Robert B. Gramacy\\ 
\vspace{-0.2cm}
{\tt bobby@statslab.cam.ac.uk}\\
\vspace{-0.2cm}
Statistical Laboratory\\
\vspace{-0.2cm}
University of Cambridge
\vspace{-0.2cm}
\and
\vspace{-0.2cm}
Herbert K. H. Lee\\
\vspace{-0.2cm}
{\tt herbie@ams.ucsc.edu}\\
\vspace{-0.2cm}
Dept of Applied Math \& Statistics\\
\vspace{-0.2cm}
University of California, Santa Cruz
\vspace{-0.2cm}
\date{}
}

\maketitle

\vspace{-0.4cm}

\begin{abstract}
\noindent
Computer experiments are often performed to allow modeling of a
response surface of a physical experiment that can be too costly or
difficult to run except using a simulator.  Running the experiment
over a dense grid can be prohibitively expensive, yet running over a
sparse design chosen in advance can result in insufficient information
in parts of the space, particularly when the surface calls for a
nonstationary model.  We propose an approach that automatically
explores the space while simultaneously fitting the response surface,
using predictive uncertainty to guide subsequent experimental runs.
The newly developed Bayesian treed Gaussian process is used as the
surrogate model, and a fully Bayesian approach allows explicit
measures of uncertainty.  We develop an adaptive sequential design
framework to cope with an asynchronous, random, agent--based
supercomputing environment, by using a hybrid approach that melds
optimal strategies from the statistics literature with flexible
strategies from the active learning literature.  The merits of this
approach are borne out in several examples, including the
motivating 
computational fluid dynamics simulation of a rocket booster.

\end{abstract}

\vspace{-0.6cm}
\section{Introduction}
\label{sec:intro}
\vspace{-0.2cm}

Many complex phenomena are difficult to investigate directly through
controlled experiments.  Instead, computer simulation is becoming a
commonplace alternative to provide insight into such phenomena
\citep{sack:welc:mitc:wynn:1989,sant:will:notz:2003}.  However, the
drive towards higher fidelity simulation continues to tax the fastest
computers, even in highly distributed computing environments.
Computational fluid dynamics (CFD) simulations in which fluid flow
phenomena are modeled are an excellent example---fluid flows over
complex surfaces may be modeled accurately but only at the cost of
supercomputer resources.  In this paper we explore the problem of
fitting a response surface for a computer model when the experiment
can be designed adaptively, i.e., online---a task to which the
Bayesian approach is particularly well--suited.  To do so, we will
combine elements from treed modeling \citep{chip:geor:mccu:2002} with
modern Bayesian surrogate modeling \citep{kennedy:ohagan:2001}, and
elements of the sequential design of computer experiments
\citep{sant:will:notz:2003} with active learning
\citep{mackay:1992,cohn:1996}.  The result is 
a fast and flexible design interface for the sequential
design of supercomputer experiments.

Consider a simulation model which defines a mapping, perhaps
non-deterministic, from parameters describing the inputs to one or
more output responses.  Without an analytic representation of this
mapping, simulations must be run for many
different input configurations in order to build up an understanding
of its possible outcomes.  
Even in extremely parallel computing environments, computational
expense of the simulation and/or high dimensional input often
prohibits the na\"{\i}ve approach of running the experiment over a
dense grid of input configurations.  More sophisticated design
strategies, such as a Latin Hypercube sample (LHS), maximin designs,
orthogonal arrays, and maximum entropy designs 
can offer an improvement over gridding.  Sequential versions of these
are better still.  However, such traditional approaches are
``stationary'' (or global, or uniform) in the sense they are based on
a metric (e.g., distance) which is measured identically throughout the
input space.  The resulting designs are sometimes called ``sparse'',
or ``space-filling''.  Maximum entropy designs are literally
stationary when based on stationary models (e.g., linear or Gaussian
Process models).  Such sparse, or stationary, design strategies are a
mismatch when the responses necessitate a nonstationary model---common
in experiments modeling physical processes, e.g., fluid dynamics---as
they cannot learn about, and thus concentrate exploration in, more
interesting or complicated regions of the input space.

For example, NASA is developing a new re-usable rocket booster called
the Langley Glide--Back Booster (LGBB).  Much of its development is
being done with computer models.  In particular, NASA is interested in
learning about the response in flight characteristics (lift, drag,
pitch, side--force, yaw, and roll) of the LGBB as a function of three
inputs (speed in Mach number, angle of attack, and side slip angle) when the
vehicle is re-entering the atmosphere.  For each input configuration
triplet, CFD simulations yield six response outputs.  
\begin{figure}[ht!]
\begin{center}
\vspace{-0.3cm}
\includegraphics[scale=0.45,angle=-90]{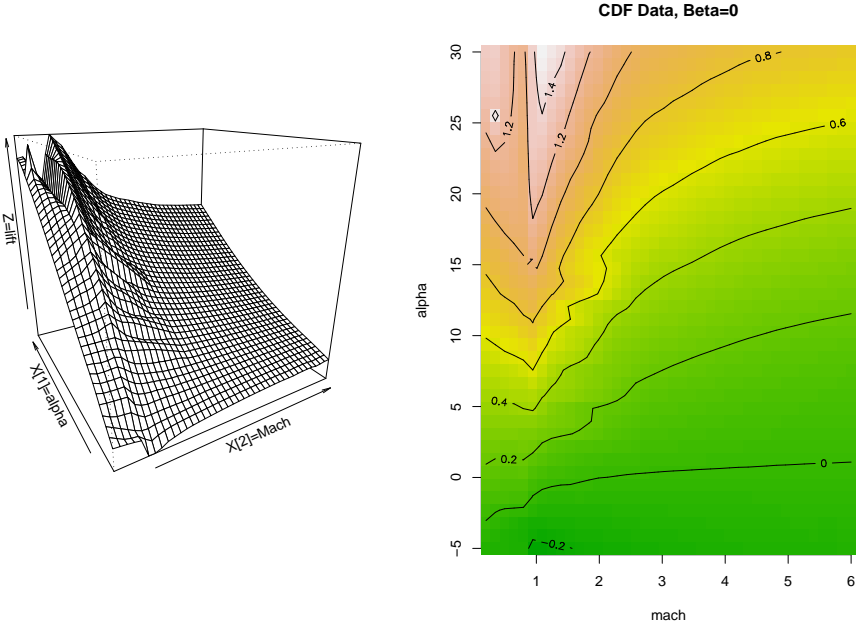}
\vspace{-0.3cm}
\caption[LGBB initial experiment]{ Lift plotted as a function of Mach
  (speed) and alpha (angle of attack) with beta (side-slip angle)
  fixed to zero.  The ridge at Mach 1 separates subsonic from
  supersonic cases.\label{f:cfdinit1}}
\end{center}
\vspace{-0.4cm}
\end{figure}
Figure~\ref{f:cfdinit1} shows the lift response as a function of speed
(Mach) and angle of attack (alpha) with the side-slip angle (beta)
fixed at zero.  The figure shows how the characteristics of subsonic
flows can be quite different from supersonic flows, as indicated by
the ridge in the response surface at Mach 1. Moreover, the transition
between subsonic and supersonic is distinctly non-linear and may even
be non-differentiable or non-continuous.  The CFD simulations involve
the iterative integration of inviscid Euler equations and are thus
computationally demanding.  Each run of the Euler solver for a given
set of parameters takes on the order of 5--20 hours on a high end
workstation.  Since simulation is expensive, there is interest in
being able to automatically and adaptively design the experiment to
learn about where response is most interesting, e.g., where
uncertainty is largest or the structure is richest, and spend
relatively more effort sampling in these areas.  However, before a
clever sequential design strategy can be devised, a nonstationary
model is needed that can capture the differences in behavior between
subsonic and supersonic physical regimes.  

The surrogate model commonly used to approximate outputs to computer
experiments is the Gaussian process (GP) \citep{sant:will:notz:2003}.
The GP is conceptually straightforward, easily accommodates prior
knowledge in the form of covariance functions, and returns estimates
of predictive confidence.  In spite of its simplicity, there are two
important disadvantages to the standard GP in this setting.  Firstly,
the computation time for inference on the GP scales poorly with the
number of data points, typically growing with the cube of the sample
size.  But most importantly, GP models are usually stationary in that
the same covariance structure is used throughout the entire input
space. In the application of high--velocity computational fluid
dynamics, where subsonic flow is quite different from supersonic flow,
this limitation is unacceptable.  Therefore, the error (standard
deviation) associated with a predicted response under a GP model does
not locally depend on any of the previously observed output responses.
The Bayesian treed GP model \citep{gra:lee:2008} was designed to
overcome these limitations.

Ideally, we would like to be able to combine the treed GP model with
classic model--based optimal design algorithms.  
However, classic design algorithms are ill--suited to partition models
and Bayesian Monte Carlo--based inference.  They are inherently serial and
thus tacitly assume a controlled and isolated computing environment.
The modern supercomputer has thousands of computing nodes, or agents,
designed to serve a multitude of diverse users.  If the design strategy
is not prepared to engage an agent as soon as it becomes available,
then that resource is either wasted or devoted to another process.  If
design is to be sequential (which it must, in order to learn about the
responses online, and adapt the model), then the interface must be
asynchronous, and any computation must execute in parallel.  There may
not be time to re-fit the model and compute the next optimal design.  So
the final ingredients in our flexible design framework are active
learning strategies from the Machine Learning literature.  Such
strategies have been used as fast, Monte Carlo--friendly, approximate
alternatives to optimal sequential design \citep{seo00}.  

Thus this paper makes two primary contributions: an integrated
sequential design strategy for a modern nonstationary model, and
methods for designing experiments in an asynchronous parallel
computing environment.  Our hybrid design strategy puts together the
treed GP model, classic sequential design, and active learning,
resulting in a highly efficient nonstationary model and sequential
design combination that balances optimality and flexibility.  
The remainder of this paper is organized as follows.  Section
\ref{sec:review} reviews the main ingredients: from conventional
optimal designs and active learning strategies with the canonical
stationary GP model, to the nonstationary treed GP model with Bayesian
model averaging and hybrid sequential design.  Section \ref{sec:as}
details our approach to the sequential design of supercomputer
experiments with the treed GP surrogate model.  Illustrative examples
are given on synthetic data in Section \ref{sec:examples}.  The
motivating example of a supercomputer experiment involving
computational fluid dynamics code for designing a re-usable launch
vehicle (LGBB) is described in Section \ref{sec:lgbb}.  Section
\ref{sec:conclude} offers some discussion and avenues for further
research.

\vspace{-0.4cm}
\section{Review}
\label{sec:review}
\vspace{-0.3cm}

Our approach to the adaptive design of supercomputer experiments
combines classic design strategies and active learning with a modern
approach to nonstationary spatial modeling.  These topics are
reviewed here.

\vspace{-0.3cm}
\subsection{Surrogate Modeling}
\label{sec:surrogate}
\vspace{-0.2cm}

In a computer experiment, the simulation output $z(\mb{x})$, 
for a particular (multivariate) input configuration
value $\mb{x}$, 
is typically modeled as a zero mean random process with covariance
$C(\mb{x}, \mb{x}') = \sigma^2 K(\mb{x}, \mb{x}')$.  The stationary
Gaussian process (GP) is a popular example of such a model, and
consequently is the canonical surrogate model used in designing computer
experiments~\citep{sack:welc:mitc:wynn:1989,sant:will:notz:2003}.

For data $D=\{\mb{x}_i^\top,z_i\}_{i=1}^N = \{\mb{X}, \mb{Z}\}$ of
$m_X$--dimensional inputs $\mb{X}$ and scalar observations $\mb{Z}$
under a GP model, the density over outputs at a new point $\mb{x}$ has
a Normal distribution
with 
\vspace{-0.3cm}
\begin{align} 
\mbox{mean} && \hat{z}(\mb{x}) 
        &= \mb{k}^\top(\mb{x}) \mb{K}^{-1} \mb{Z}, 
        \; \mbox{ and} \nonumber \\ 
\mbox{variance} && \hat{\sigma}^2(\mb{x}) &= \sigma^2
        [K(\mb{x},\mb{x}) - \mb{k}^\top(\mb{x})
        \mb{K}^{-1} \mb{k}(\mb{x})] \label{eq:pvar} 
\end{align} 
where $\mb{k}^\top(\mb{x})$ is the $N$--vector whose $i^{\mbox{\tiny
    th}}$ component is $K(\mb{x},\mb{x}_i)$, and $\mb{K}$ is the
$N\times N$ matrix with $i,j$ element $K(\mb{x}_i, \mb{x}_j)$.  It is
important to note that the uncertainty, $\hat{\sigma}^2(\mb{x})$,
associated with the prediction has no direct dependence on the nearby
observed simulation outputs $\mb{Z}$; because of the assumption of
stationarity, all response points contribute to the estimation of the
local error through their influence on the correlation function
$K(\cdot,\cdot)$, and the induced correlation matrix
$\mb{K}_{i,j}=K(\mb{x}_i,\mb{x}_j)$.  We follow Gramacy and Lee (2008)
in specifying that $K(\cdot,\cdot)$ have the form \vspace{-0.2cm}
\begin{equation} 
K(\mb{x}_j, \mb{x}_k|g) =
        K^*(\mb{x}_j, \mb{x}_k) + {g} \delta_{j,k}, \label{eq:cor}
\end{equation} 
where $\delta_{\cdot,\cdot}$ is the Kronecker delta function, and
$K^*$ is a {\em true} correlation function.  The $g$ term, referred to
as the {\em nugget}, is positive $(g>0)$ and provides a mechanism for
introducing measurement error into the stochastic process.  It arises
when considering a model of the form $z(\mb{x}) = w(\mb{x}) +
\eta(\mb{x})$ where $w(\mb{x})$ is the random process with covariance
$C$, and $\eta(\cdot)$ is independent Gaussian noise.  Valid
correlation functions $K^*(\cdot,\cdot)$ are usually generated as a
member of a parametric family, such as the 
separable power or Mat\'{e}rn families.  A general reference for
families of correlation functions $K^*$ is provided by
\citet{abraham:1997}.  Hereafter we use the separable power family,
\begin{equation} 
        K^*(\mb{x}_j, \mb{x}_k|\mb{d}) = 
        \label{eq:cor_d} \exp\left\{ - \sum_{i=1}^{m_X}
        \frac{|x_{ij} - x_{ik}|^{p_0}}{d_{i}}\right\},
\end{equation} 
which is a standard choice in modeling computer experiments
\citep{sant:will:notz:2003}.  We fix $p_0 = 2$ and infer the range
parameters $\{d_i\}_{i=1}^{m_X}$ as part of our estimation procedure.

While many authors
\citep[e.g.,][]{sant:will:notz:2003,sack:welc:mitc:wynn:1989}
deliberately omit the nugget parameter on the grounds that computer
experiments are deterministic, we have found it more helpful to
include a nugget.  Most importantly, we found that the LGBB simulator
was only theoretically deterministic, but not necessarily so in
practice.  Researchers at NASA explained to us that their numerical
CFD solvers are typically started with random initial values, and
involve forced random restarts when diagnostics indicate that
convergence is poor.  Furthermore, due to the sometimes chaotic
behavior of the systems, input configurations arbitrarily close to one
another can fail to achieve the same estimated convergence, even after
satisfying the same stopping criterion.  Thus a conventional GP model
without a small--distance noise process (nugget) can be a mismatch to
such potentially non-smooth data.  As a secondary concern, numerical
stability in decomposing covariance matrices can be improved by using
a small nugget term \citep{neal:1997}.

\vspace{-0.3cm}
\subsubsection{Treed Gaussian process model}
\label{sec:tgp}
\vspace{-0.2cm}

Because of concerns about the inadequacy of the stationarity
assumption, we propose a surrogate model that is new in the realm of
sequential design of experiments: the Bayesian
treed Gaussian process (treed GP) model \citep{gra:lee:2008}.  The
treed GP model extends the Bayesian treed linear model by using a GP model with
linear trend independently within each region, instead of constant
\citep{chip:geor:mccu:1998,deni:mall:smit:1998} or linear
\citep{chip:geor:mccu:2002} models in the partitions.  A process prior
\citep{chip:geor:mccu:1998} is placed on the tree $\mathcal{T}$, and
conditional on $\mathcal{T}$, parameters for $R$ independent GPs in
regions $\{r_\nu\}_{\nu=1}^R$ are specified via a hierarchical
generative model: 
\vspace{-0.3cm}
\begin{align} 
\mb{Z}_\nu | \bm{\beta}_\nu, \sigma^2_\nu, \mb{K}_\nu &\sim 
N_{n_\nu}(\mb{\mb{F}}_\nu \bm{\beta}_\nu, \sigma^2_\nu \mb{K}_\nu)  & 
\bm{\beta}_0 &\sim N_{m}(\bm{\mu}, \mb{B}) &
\sigma^2_\nu &\sim IG(\alpha_\sigma/2, q_\sigma/2) \label{eq:model} \\
\bm{\beta}_\nu | \sigma^2_\nu, \tau^2_\nu, \mb{W}, 
\bm{\beta}_0 &\sim N_{m}(\bm{\beta}_0,\sigma^2_\nu \tau^2_\nu \mb{W}) &
\mb{W}^{-1} &\sim W((\rho \mb{V})^{-1}, \rho) &
\tau^2_\nu &\sim IG(\alpha_\tau/2, q_\tau/2) \nonumber
\end{align} 
where $\mb{F}_\nu = (\mb{1}, \mb{X}_\nu)$, $\mb{W}$ is a $m \times
m$ matrix, and $m=m_X + 1$.  $N$, $IG$, and $W$ are the (Multivariate)
Normal, Inverse--Gamma, and Wishart distributions, respectively.
$\mb{K}_\nu$ is the separable power family covariance matrix with a
nugget, as in~(\ref{eq:cor}--\ref{eq:cor_d}).  The data
$\{\mb{X},\mb{Z}\}_\nu$ in region $r_\nu$ are used to estimate the
parameters $\bm{\theta}_\nu = \{{\bm \beta}, \sigma^2, \bm{K},
\tau^2\}_\nu$ of the model active in the region.  Parameters to the
hierarchical priors depend only on $\{\bm{\theta}_\nu\}_{\nu=1}^R$.
Samples from the posterior distribution are gathered using Markov
chain Monte Carlo (MCMC).  All parameters can be sampled using Gibbs
steps, except for the covariance structure, whose parameters can be
sampled via Metropolis--Hastings.

The predicted value of $Z(\mb{x}\in r_{\nu})$ is normally distributed
with mean and variance
\vspace{-0.3cm}
\begin{align} 
 \hat{z}(\mb{x}) &= E(Z(\mb{x}) | 
        \; \mbox{data}, \mb{x}\in r_\nu ) \; = \;
 \mb{f}^\top(\mb{x}) \tilde{\bm{\beta}}_\nu +
        \mb{k}_\nu(\mb{x})^\top \mb{K}_\nu^{-1}(\mb{Z}_\nu -
        \mb{F}_\nu\tilde{\bm{\beta}}_\nu), \label{eq:predmean} \\ 
 \hat{\sigma}^2(\mb{x}) &= 
        \mbox{Var}(Z(\mb{x}) | \;\mbox{data}, \mb{x}\in r_\nu ) 
  \; = \; \sigma_\nu^2 [\kappa(\mb{x}, \mb{x}) - 
        \mb{q}_\nu^\top(\mb{x})\mb{C}_\nu^{-1} \mb{q}_\nu(\mb{x})],
        \label{eq:predvar} 
\end{align}
\vspace{-1.5cm}
\begin{align}
\mbox{where} &&
\mb{C}_\nu^{-1} &= (\mb{K}_\nu + 
        \tau_\nu^2\mb{F}_\nu \mb{W} \mb{F}_\nu^\top)^{-1} &
\mb{q}_\nu(\mb{x}) &= 
        \mb{k}_\nu(\mb{x}) + \tau_\nu^2\mb{F}_\nu \mb{W}_\nu 
        \mb{f}(\mb{x}) \label{eq:auxpred} \\ 
&& \kappa(\mb{x},\mb{y}) &= K_\nu(\mb{x},\mb{y}) 
        + \tau_\nu^2\mb{f}^\top(\mb{x}) \mb{W} \mb{f}(\mb{y}) \nonumber
\end{align} 
with $\mb{f}^\top(\mb{x}) = (1, \mb{x}^\top)$, and
$\mb{k}_\nu(\mb{x})$ a $n_\nu-$vector with $\mb{k}_{\nu,j}(\mb{x})=
K_\nu(\mb{x}, \mb{x}_j)$, for all $\mb{x}_j \in \mb{X}_\nu$.  The
global process is nonstationary because of the tree ($\mathcal{T}$)
and thus $\hat{\sigma}^2(\mb{x})$ in~(\ref{eq:predvar}) is
region--specific.  The predictive surface can be discontinuous across
the partition boundaries of a particular tree $\mathcal{T}$.  However,
in the aggregate of samples collected from the joint posterior
distribution of $\{\mathcal{T}, \bm{\theta}\}$, the mean tends to
smooth out near likely partition boundaries as the tree operations
{\em grow, prune, change, swap}, and {\em rotate} integrate over trees
and GPs with larger posterior probability \citep{gra:lee:2008}.
Uncertainty in the posterior for $\mathcal{T}$ translates into higher
posterior predictive uncertainty near region boundaries.  When the
data actually indicate a non-smooth process, e.g., as in the LGBB
experiment in Section~\ref{sec:lgbb}, the treed GP retains the
capability to model discontinuities.

The Bayesian treed linear model of \cite{chip:geor:mccu:2002} is
implemented as a special case of the treed GP model, called the treed
GP LLM (short for: ``with jumps to the Limiting Linear Model'').
Detection of linearity in the response surface is facilitated on a
per-dimension basis via the introduction of $m_X$
indicator--parameters $\mb{b}_\nu$, in each region $r_\nu=1,\dots,R$,
which are given a prior conditional on the range parameter(s) to
$K_\nu(\cdot,\cdot)$. The boolean $b_{\nu i}$ determines whether the
GP or its LLM governs the marginal process in the $i^{\mbox{\tiny
    th}}$ dimension of region $r_\nu$.  The result, through Bayesian
model averaging, is an adaptively semiparametric nonstationary
regression model which can be faster, more parsimonious, and
numerically stable \citep{gra:lee:2008b}.  Empirical evidence suggests
that many computer experiments involve responses which are either
linear in most of the input dimensions, or entirely linear in a subset
of the input domain [see Section~\ref{sec:lgbb}].  Thus the treed GP
LLM is particularly well--suited to be a surrogate model for computer
experiments.

Compared to other approaches to nonstationary modeling, including
using spatial deformations
\citep{samp:gutt:1992,schmidt:2003} and process convolutions
\citep{higd:swal:kern:1999,Paci:2003}, the treed GP LLM approach
yields an extremely fast implementation of nonstationary GPs,
providing a divide--and--conquer approach to spatial modeling.
Although the method is especially well--suited to axis--aligned
nonstationarity, which is common in computer experiments, it has been
found to compare favorably in situations when the nature of the
nonstationarity is more general \citep{gra:lee:2008,gra:lee:2008b}.
For example, in Section \ref{sec:as:exp} we consider a dataset where
the treed GP is quite appropriate even though the correlation
structure varies radially, and moreover, which is well--fit by a
stationary model for small designs.  In Section \ref{sec:as:sixd} we
consider a high dimensional dataset where the nature of the
nonstationarity is unknown.  Software implementing the treed GP LLM
model and all of its special cases (e.g., stationary GP, CART \& the
treed linear model, linear model, etc.)  is available as an {\sf R}
package~\citep{cran:R}, and can be obtained from CRAN:
\begin{center}
\verb!http://www.cran.r-project.org/web/packages/tgp/index.html!.
\end{center}
The package implements a family of default prior specifications, i.e.,
settings for the constants in (\ref{eq:model}).  In this paper we use
these defaults unless otherwise noted.  For more details see the {\tt
  tgp} documentation \citep{package:tgp} and tutorial
\citep{Gramacy:2007:JSSOBK:v19i09}.

\vspace{-0.2cm}
\subsection{Sequential design of experiments}
\label{sec:intro:as}
\vspace{-0.1cm}

In the statistics community, the traditional approach to sequential
data solicitation is called {\em (Sequential) Design of Experiments}
(DOE) or {\em Sequential Design and Analysis of Computer Experiments}
(SDACE) when applied to computer simulations
\citep{sack:welc:mitc:wynn:1989,currin:1991,welch:1992,sant:will:notz:2003}.
Depending on whether the goal of the experiment is inference or
prediction, as described by a choice of utility, different algorithms
for obtaining optimal designs can be derived.  For example, one can
choose the Kullback--Leibler distance between the posterior and prior
distributions as a utility.  For Gaussian process models with
correlation matrix $\mb{K}$, this is equivalent to maximizing
det$(\mb{K})$.  Subsequently chosen input configurations are called
maximum entropy designs \citep[e.g.,][Chapter
6]{shewry:wynn:1987,sant:will:notz:2003}.
An excellent review of Bayesian approaches to DOE is
provided by \citet{chaloner:1995}.

Finding optimal designs can be computationally intensive, especially
for stationary GP surrogate models, because the algorithms usually
involve repeated decompositions of large covariance matrices.
Determinant--space, for example, can have many local maxima which can
be sought--after via stochastic search, i.e., simulated annealing,
genetic algorithms \citep{hamada:2001}, etc.  A parametric family is
assumed, either with fixed parameter values, or a preliminary analysis
is used to find maximum likelihood estimates for its parameters, which
are then treated as ``known''.  In a sequential design, parameters
estimated from previous designs can be used, whereas a Bayesian
decision theoretic approach may ``choose'' a parameterization and
optimal design jointly~\citep{muller:sanso:dei:2004}.  In all of these
approaches, it is important to note that optimality is only with
respect to the assumed parametric form.  Should this form not be known
a priori, as is often the case in practice, then the resulting designs
could be far from optimal.

Other nonparametric approaches used in the statistics literature
include space filling designs, e.g.,~maximin distance designs and
LHS~\citep{mcka:cono:beck:1979,sant:will:notz:2003}.  Computing
maximin distance designs can also be computationally intensive,
whereas LHSs are easy to compute and result in well--spaced
configurations relative to random sampling, though there are some
degenerate cases, such as diagonal LHSs~\citep{sant:will:notz:2003}.
LHSs can also be less advantageous in a sequential sampling
environment since there is no mechanism to ensure that the
configurations will be well--spaced relative to previously sampled
(fixed) locations.  Maximum entropy designs, and maximin designs, may
be more computationally demanding, but they are easily converted into
sequential design methods by simply fixing the locations of samples
whose response has already been obtained, and then optimizing only
over new sample locations.

\vspace{-0.2cm}
\subsubsection{An active learning approach sequential experimental
  design}
\label{sec:intro:hybrid}
\vspace{-0.2cm}

In the world of Machine learning, design of experiments would (loosely)
fall under the blanket of a research focus called {\em active
  learning}.  In the
literature~\citep{angluin:1988,atlas:cohn:2003},
active learning, or equivalently {\em query learning} or {\em
  selective sampling}, refers to the situation where a learning
algorithm has some, perhaps limited, control over the inputs it trains
on.  There are essentially two active learning approaches to DOE
using the GP.  The first approach tries to maximize the information
gained about model parameters by selecting from a set of candidates
$\tilde{\mb{X}}$, the location $\tilde{\mb{x}}\in\tilde{\mb{X}}$ which
has the greatest standard deviation in predicted output.  This
approach, called ALM for Active Learning--MacKay, has been shown to
approximate maximum expected information designs \citep{mackay:1992}.
\citet{klei:vanb:2004} take a similar approach.

An alternative algorithm, called ALC for Active Learning--Cohn, is to
select $\tilde{\mb{x}} \in \tilde{\mb{X}}$ maximizing the expected
reduction in squared error averaged over the input space
\citep{cohn:1996}.
Using the notation from (\ref{eq:pvar}) for stationary GPs, and
supposing that the location $\tilde{\mb{x}}$ is added into the design,
a global reduction in predictive variance can be obtained by averaging
over other locations $\mb{y}$: \vspace{-0.2cm}
\begin{align}
\Delta \hat{\sigma}^2 (\tilde{\mb{x}}) &= 
        \int_\mb{y} \Delta\hat{\sigma}^2_{\tilde{\mb{x}}} (\mb{y}) 
        = \int_\mb{y} \hat{\sigma}^2(\mb{y}) - \hat{\sigma}^2_{\tilde{\mb{x}}}
        (\mb{y}) 
        = \int_\mb{y} \frac{\sigma^2 \left[ 
        \mb{k}^\top(\mb{y}) \mb{K}_N^{-1} \mb{k}(\tilde{\mb{x}}) 
        - K(\tilde{\mb{x}}, \mb{y}) \right]^2}
        {K(\tilde{\mb{x}}, \tilde{\mb{x}}) 
        - \mb{k}(\tilde{\mb{x}})^\top\mb{K}_N^{-1}\mb{k}(\tilde{\mb{x}})}.
\label{eq:ialc}
\end{align}
In practice the integral is replaced by a sum over a grid of locations
$\tilde{\mb{Y}}$, typically with $\tilde{\mb{Y}} = \tilde{\mb{X}}$,
and the parameterization to the model, i.e.,~$K(\cdot, \cdot)$ and
$\sigma^2$, is assumed known in advance.  \citet{seo00} provide a
comparison for stationary GPs between ALC and ALM.

\vspace{-0.2cm}
\subsubsection{Other  approaches to designing computer experiments}
\label{sec:intro:otherdoe}
\vspace{-0.2cm}

Bayesian and non-Bayesian approaches to surrogate modeling and design
for computer experiments abound
\citep{sack:welc:mitc:wynn:1989,currin:1991,welch:1992,bates:1996,sebast:wynn:2000,kennedy:ohagan:2000,kennedy:ohagan:2001}.
A recent approach, which bears some similarity to ours, uses
stationary GPs and a so--called {\em spatial aggregate language} to
aid in an {\em active data mining} of the input space of the
experiment \citep{rama:2005}.  Our use of nonstationary surrogate
models within a highly distributed supercomputer architecture
distinguishes our work from the methods described in those papers,
and yields a more dynamic approach to sequential design.

\vspace{-0.4cm}
\section{Adaptive sequential design}
\label{sec:as}
\vspace{-0.3cm}

Much of the current work in large scale computer models starts by
evaluating the model over a hand crafted set of input configurations,
such as a full grid or some reduced design.  After the initial set has
been run, a human may identify interesting regions and perform
additional runs if desired.  We are concerned with automating this
process, 
based on local estimates of uncertainty
that can provided by the nonstationary treed GP LLM surrogate
model.\footnote{We shall drop the LLM tag in what follows, and consider
  it implied by the label treed GP.}

\vspace{-0.3cm}
\subsection{Asynchronous distributed supercomputing}
\label{sec:as:comp}
\vspace{-0.2cm}

High fidelity supercomputer experiments are usually run on clusters of
independent computing agents, or processors.  A {\tt Beowulf} cluster
is a good example.  At any given time, each agent is working on a
single input configuration.  Multiple agents allow several input
configurations to be run in parallel.  Simulations for new
configurations begin when an agent finishes execution and becomes
available.  Therefore, simulations may start and finish at different,
perhaps even random, times.  The cluster is managed asynchronously by
a master controller {\em (emcee)} program that gathers responses from
finished simulations, and supplies free agents with new input
configurations.
\begin{figure}
\begin{center}
\vspace{-0.2cm}
\includegraphics[scale=0.6]{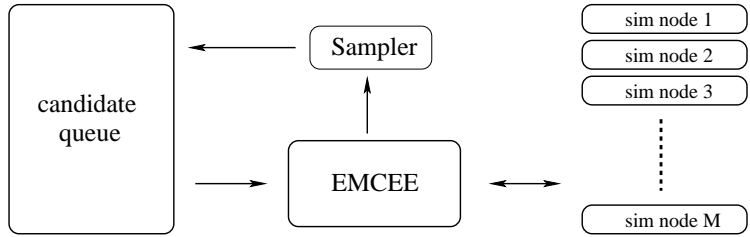}
\vspace{-0.1cm}
\caption[Supercomputer {\em emcee} interacts with adaptive sampler]
{{\em Emcee} program gives finished simulations to the sampler (which
  populates the queue based on a surrogate model) and gets new ones
  from the queue. \label{f:emcee}}
\end{center}
\vspace{-0.2cm}
\end{figure}
The goal is to have the {\em emcee} program interact with a
nonstationary modeling and sequential design program that maintains a
queue of well--chosen candidates, and to which it provides finished
responses as they become available, so that the surrogate model can be
updated [see Figure~\ref{f:emcee}].

\vspace{-0.3cm}
\subsection{Adaptive sequential DOE via active learning}
\label{sec:as:sdoe}
\vspace{-0.2cm}

In the statistics community, there are a number of established
methodologies for (sequentially) designing experiments [see Section
\ref{sec:intro:as}].  However, some classic criticisms for traditional
DOE approaches precluded such an approach here.  The primary issue is
that ``optimally'' chosen design points are usually along the boundary
of the region, where measurement error can be severe, responses can be
difficult to elicit, and model checking is often not
feasible~\citep{chaloner:1995}.
Furthermore, boundary points are only optimal when the model is known
precisely.  For example, in one--dimensional linear regression, putting
half the points at one boundary and half at the other is only optimal
if the true model is linear; if it turns out that the truth may be
quadratic, then forcing all points to the boundaries is highly
suboptimal.  Similarly here, where we do not know the full form of the
model in advance, it is important to favor internal points so that the
model (including the partitions) can be learned correctly.  
Other drawbacks to the traditional DOE approach include speed, the
difficulty inherent in using Monte Carlo to estimate the surrogate
model, lack of support for partition models, and the desire to design
for an asynchronous {\em emcee} interface where responses and
computing nodes become available at random times.

Instead, we take a two-stage (hybrid) approach that combines standard
DOE with methods from the active learning literature.  The first stage
is to use optimal sequential designs from the DOE literature, such as
maximum entropy, maximin designs, or LHS, as {\em candidates} for
future sampling.  This ensures that candidates for future sampling are
well--spaced out relative to themselves, and to the already sampled
locations.  In the second stage, the treed GP surrogate model can
provide Monte Carlo estimates of region--specific model uncertainty,
via the ALM or ALC algorithm, which can be used to populate, and
sequence, the candidate queue used by the {\em emcee} [see
Figure~\ref{f:emcee}].  This ensures that the most informative of the
optimally spaced candidates can be first in line for simulation when
agents become available.

\vspace{-0.3cm}
\subsection{ALM and ALC algorithms}
\label{sec:as:hybrid}
\vspace{-0.2cm}

Given a set of candidate input configurations $\tilde{\mb{X}}$,
Section~\ref{sec:intro:hybrid} introduced two active learning criteria
for choosing amongst---or ordering---them based on the
posterior predictive distribution.  ALM chooses the
$\tilde{\mb{x}}\in\tilde{\mb{X}}$ with the greatest standard deviation
in predicted output~\citep{mackay:1992}.  MCMC posterior predictive
samples provide a convenient estimate of location--specific variance,
namely the width of predictive quantiles.

Alternatively, ALC selects the $\tilde{\mb{x}}$ that maximizes the
expected reduction in squared error averaged over the input space
\citep{cohn:1996}.  Conditioning on $\mathcal{T}$, the reduction in
variance at a point $\mb{y}\in r_\nu$, given that the location
$\tilde{\mb{x}}\in\tilde{\mb{X}}_\nu$ is added into the data, is
defined as (region subscripts suppressed): \vspace{-0.3cm}
\begin{align*} 
\vspace{-0.3cm}
\Delta \hat{\sigma}^2_{\tilde{\mb{x}}} (\mb{y}) &= 
        \hat{\sigma}^2(\mb{y}) - \hat{\sigma}^2_{\tilde{\mb{x}}} (\mb{y}) &&&
\mbox{where} && 
\hat{\sigma}^2(\mb{y}) &= 
        \sigma^2[\kappa(\mb{y}, \mb{y}) - \mb{q}_N^\top(\mb{y}) \mb{C}_N^{-1}
        \mb{q}_N^\top(\mb{y})], \\ 
&&&& \mbox{and} &&
\hat{\sigma}^2_{\tilde{\mb{x}}}(\mb{y})
        &= \sigma^2[\kappa(\mb{y}, \mb{y}) - \mb{q}_{N+1}^\top(\mb{y})
        \mb{C}_{N+1}^{-1}
        \mb{q}_{N+1}(\mb{y})]
\end{align*} 
using notation for the GP predictive variance for region $r_\nu$ given
in~(\ref{eq:predvar}).  Note that the $N+1^{\mbox{\tiny st}}$
component of $\mb{q}_{N+1}(\mb{y})$, and the corresponding column and
row of $\mb{C}_{N+1}$, are a function of $\tilde{\mb{x}}$.  The
partition inverse equations \citep{barnett:1979}, for a covariance
matrix $\mb{C}_{N+1}$ in terms of $\mb{C}_N$, yield:
\begin{equation}
\Delta \hat{\sigma}^2_{\tilde{\mb{x}}} (\mb{y}) =
        \frac{\sigma^2 \left[ \mb{q}_N^\top(\mb{y}) \mb{C}_N^{-1} 
        \mb{q}_N(\tilde{\mb{x}}) -
        \kappa(\tilde{\mb{x}}, \mb{y}) \right]^2} {\kappa(\tilde{\mb{x}}, \tilde{\mb{x}}) -
        \mb{q}_N^\top(\tilde{\mb{x}})\mb{C}_N^{-1}\mb{q}_N(\tilde{\mb{x}})}.  \label{e:gpalc}
\end{equation} 
The details of this derivation are included in Appendix
\ref{sec:a:alcgp}.  For $\mb{y}$ and $\tilde{\mb{x}}$ not in the same
region $r_\nu$, let $\Delta\sigma^2_{\tilde{\mb{x}}}(\mb{y}) = 0$.
Rather than integrating, as in (\ref{eq:ialc}), the reduction in
predictive variance that would be obtained by adding $\tilde{\mb{x}}$
into the dataset is calculated in practice by averaging over a grid
or candidate set of $\mb{y}\in\mb{Y}$: 
\begin{equation} 
\Delta \sigma^2(\tilde{\mb{x}}) =
|\mb{Y}|^{-1} 
\sum_{\mb{y} \in\mb{Y}} \Delta \hat{\sigma}^2_{\tilde{\mb{x}}} (\mb{y}) 
\label{e:alcmean}
\end{equation} 
which can be approximated using MCMC methods.  Compared to ALM,
adaptive samples under ALC are less heavily concentrated near the
boundaries of the partitions.  Both provide a ranking of a set of
candidate locations $\tilde{\mb{x}}\in\tilde{\mb{X}}$.  Computational
demands are in $O(|\tilde{\mb{X}}|)$ for ALM, and
$O(|\tilde{\mb{X}}||\mb{Y}|)$ for ALC.  
\citet{mcka:cono:beck:1979} provide a comparison between ALM, and LHS,
on computer code data.  Seo et al.~(2000) provide comparisons between
ALC and ALM using standard GPs, taking $\mb{Y} = \tilde{\mb{X}}$ to be
the full set of un-sampled locations in a pre-specified dense uniform
grid.  In both papers, the model is assumed known in advance.

However, that last assumption, that the model is known {\em a priori}
is at loggerheads with sequential design---if the model were already
known then why design sequentially?  In the treed GP application of
ALC, the model is not assumed known {\em a priori}.  Instead, Bayesian
MCMC posterior inference on $\{\mathcal{T}, \bm{\theta}\}$ is
performed, and then samples from
$\Delta\sigma^2_{\tilde{\mb{x}}}(\mb{y})$ are taken conditional on
samples from $\{\mathcal{T}, \bm{\theta}\}$.  To mitigate the
(possibly enormous) expense of sampling
$\Delta\sigma^2_{\tilde{\mb{x}}}(\mb{y})$ via MCMC on a dense
high--dimensional grid (with $\mb{Y} = \tilde{\mb{X}}$), a smaller and
more cleverly--chosen set of candidates can come from the sequential
treed maximum entropy design, described in the following subsection.
The idea is to sequentially select candidates which are well--spaced
relative both to themselves and to the already sampled configurations,
in order to encourage
exploration. 

Applying the ALC algorithm under the limiting linear model (LLM) is
computationally less intensive compared to ALC under a full GP.
Starting with the predictive variance given in~(\ref{eq:auxpred}),
the expected reduction in variance under the linear model 
is given in~(\ref{e:llmalc}), below, and averaging over
$\mb{y}$ proceeds as in~(\ref{e:alcmean}), above.
\vspace{-0.1cm}
\begin{equation}
  \Delta \hat{\sigma}^2_{\tilde{\mb{x}}} (\mb{y}) = \frac{ 
    \sigma^2 [\mb{f}^\top(\mb{y}) \mb{V}_{\tilde{\beta}_N} \mb{f}(\tilde{\mb{x}})]^2}
  {1+g + \mb{f}^\top(\tilde{\mb{x}}) \mb{V}_{\tilde{\beta}_N} \mb{f}(\tilde{\mb{x}})}
  \label{e:llmalc}
\end{equation}
Appendix \ref{sec:a:alclm} contains details of the derivation.  The $m
\times m$ matrix $\mb{V}_{\tilde{\beta}_N}$ is the posterior variance
of ${\bm \beta}$ based on the $N$ data points in the current
design. Since only an $O(m^3)$ inverse operation is required,
Eq.~(\ref{e:llmalc}) is preferred over replacing $\mb{K}$ with the
$N\times N$ matrix $\mb{I}(1+g)$ in (\ref{e:gpalc}), which requires an
$O(N^3)$ inverse.

\vspace{-0.1cm}
\subsection{Choosing candidates}
\label{sec:as:cands}
\vspace{-0.1cm}

We have already discussed how a large, i.e., densely gridded,
candidate set $\tilde{\mb{X}}$ can make for computationally expensive
ALM and (especially) ALC calculations.  In an asynchronous parallel
environment, there is another reason why candidate designs should not
be too dense.  Suppose we are using the ALM algorithm, and we estimate
the uncertainty to be highest in a particular region of the space.  If
two candidates are close to each other in this region, then they will
have the highest and second--highest priority, and the emcee could
send both of them to agents.  However, if we knew we were going to
send off two runs, we generally would not want to pick those two right
next to each other, but would want to pick two points from different
parts of the space.  If each design point could be picked
sequentially, then the candidate spacing is not an issue, because the
model can be re-fit and the points re-ordered between runs.  In the
reality of an asynchronous parallel environment, there may not be time
to re-fit the model before the emcee needs an additional run
configuration to send to another agent.  Thus there is a real need for
well--spaced candidates.

A sequential maximum entropy design \citep[][Chapter
6]{shewry:wynn:1987,sack:welc:mitc:wynn:1989,currin:1991,welch:1992,sant:will:notz:2003}
may seem like a reasonable approach because it encourages exploration.
But traditional maximum entropy designs are based on a known
parameterization of a single GP model, and are thus not well--suited
to MCMC based treed partition models wherein ``closeness'' is not
measured homogeneously throughout the input space.  Furthermore, a
maximum entropy design may not choose candidates in the
``interesting'' part of the input space because sampling is high there
already.  E.g., in the rocket booster application we want to continue
to sample close to Mach one because we need many more points to
understand the function where it is changing quickly.  Another
disadvantage to maximum entropy designs is computational, requiring
repeated decompositions of large covariance matrices.

One possible solution to both computational and nonstationary modeling
issues is to use what we call a (sequential) treed maximum entropy
design.  That is, a separate sequential maximum entropy design can be
obtained in each of the partitions depicted by the maximum {\em a
  posteriori} (MAP) tree $\hat{\mathcal{T}}$.  The number of
candidates selected from each region,
$\{\hat{r}_\nu\}_{\nu=1}^{\hat{R}}$ of $\hat{\mathcal{T}}$, can be
proportional to the volume or the number of grid locations in the
region.  MAP parameters $\hat{\bm{\theta}}_\nu|\hat{\mathcal{T}}$ can
be used in creating the candidate design, or ``neutral'' or
``exploration encouraging'' parameters can be used instead.
Separating design from inference by using custom parameterizations in
design steps, rather than inferred ones, is a common practice in the
SDACE community \citep{sant:will:notz:2003}.  Small range parameters,
for learning about the wiggliness of the response, and a modest nugget
parameter for numerical stability, tend to work well together.

Since optimal design is only used to select candidates, and is not the
final step in adaptively choosing samples, employing a high-powered
search algorithm (e.g., a genetic algorithm) is unnecessary.  Finding
a local maximum is generally sufficient to get well--spaced
candidates.  We use a simple stochastic ascent algorithm\footnote{For
  example, we find that the following works well for constructing a
  set of candidates $\tilde{\mb{X}}$ of size $|\tilde{\mb{X}}|=N'$.
  Construct a LHS $\mb{L}$ of size $10N'$, and initialize
  $\tilde{\mb{X}}$ to be a random subsample of $\mb{L}$ of size $N'$,
  without replacement. Then randomly propose to swap a single element
  of $\tilde{\mb{X}}$ with one from $\mb{L}\setminus\tilde{\mb{X}}$
  and accept only upon an observed increase in
  det$(\mb{K}([\mb{X},\tilde{\mb{X}}]))$.  Repeat until the acceptance
  rate is low, possibly with reference to $N'$.}  in each of the
$\hat{R}$ regions $\{r_\nu\}_{\nu=1}^{\hat{R}}$ of $\hat{\mathcal{T}}$
to find local maxima without calculating too many determinants.  The
$\hat{R}$ search algorithms can be run in parallel, and typically
invert matrices much smaller than $N\times N$.

\begin{figure}[ht!]
\begin{center}
\vspace{-0.2cm}
\includegraphics[angle=-90,scale=0.3]{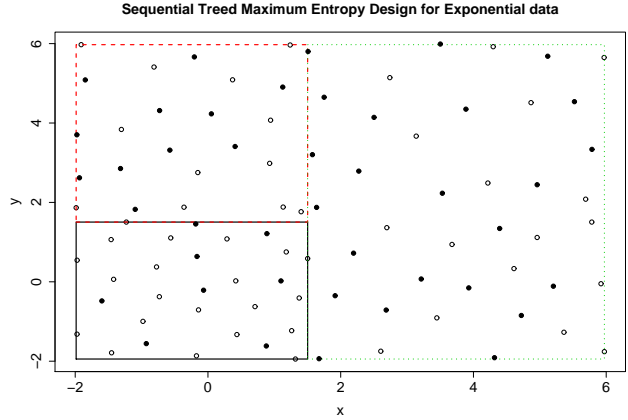}
\caption{Example of a treed maximum entropy design in 2-d.  {\em Open
    Circles} represent previously sampled locations. {\em Solid dots}
  are the candidate design based on $\hat{\mathcal{T}}$, also shown.
  \label{f:treeddopt}}
\end{center}
\vspace{-0.4cm}
\end{figure}

Figure \ref{f:treeddopt} shows an example sequential treed maximum
entropy design for the 2-d Exponential data [Section
\ref{sec:as:exp}], found by simple stochastic search. Input
configurations are sub-sampled from a LHS of size 400, and the chosen
candidate design is of size $\sim$40 (i.e., $\lceil 10\% \rceil$).
Dots in the figure represent the chosen locations of the new candidate
design $\tilde{\mb{X}}$ relative to the existing sampled locations
$\mb{X}$ (circles).
Candidates are reasonably spaced--out relative to one another, and to
existing inputs, except possibly near partition boundaries.  There are
roughly the same number of candidates in each quadrant, despite the
fact that the density of samples (circles) in the first quadrant is
almost two-times that of the others.  
A classical (non-treed) maximum entropy design would have chosen fewer
points in the first quadrant, where all the action is, in order to
equalize the density relative to the other three quadrants.

\vspace{-0.3cm}
\subsection{Implementation methodology}
\label{sec:as:methods}
\vspace{-0.2cm}

{\em Bayesian adaptive sampling} (BAS) proceeds in trials.  Suppose
that $N$ samples and their responses have been gathered in previous
trials, or from a small initial design before the first trial.  In the
current trial, a treed GP model is estimated for data
$\{\mb{x}^\top_i, z_i\}_{i=1}^N$.  Samples are gathered in accordance
with ALM or ALC conditional on $\{\bm{\theta}, \mathcal{T}\}$, at
candidate locations $\tilde{\mb{X}}$ chosen to follow a sequential
treed maximum entropy design, using the MAP tree obtained in the
previous trial.  The candidate queue is then populated with a sorted
list of candidates.  BAS gathers finished and running input
configurations from the {\em emcee} and adds them into the design.
Predictive mean estimates are used as surrogate responses for
unfinished (running) configurations until the true response is
available.  New trials start with fresh candidates.

An artificial clustered simulation environment, with a fixed number of
agents, was developed in order to simulate the parallel and
asynchronous evaluation of input configurations, whose responses
finish at random times.
It was implemented {\tt Perl} and was designed to mimic, and interface
with, the {\tt Perl} modules at NASA which drive their experimental
design software.  Experiments on synthetic data, in the next section,
will use this artificial environment.  The LGBB experiment in Section
\ref{sec:lgbb} uses the real {\tt Perl} modules to submit jobs to the
real NASA supercomputer.  Multi-dimensional responses, as in the LGBB
experiment, are treated as independent.  That is, each response has
its own treed GP surrogate model, $m_Z$ surrogates total for an
$m_Z$--dimensional response.  Uncertainty estimates (via ALM or ALC)
are normalized and pooled across the models for each response in order
to develop a single (sequential) design for the entire process.
Treating highly correlated physical measurements as independent is a
crude approach.  However, it still affords remarkable results, and
allows the use of the {\tt PThreads} parallel computing library to get
a highly parallel implementation and take advantage of multi-core
processors that are becoming commonplace. Coupled with the
producer/consumer model for parallelizing prediction and estimation
\citep{gra:lee:2008}, a factor of $2m_Z$ speedup for $2m_Z$ processors
can be obtained.\footnote{For more information on the parallel
  implementation, please see Appendix C.2 of
  \citet{Gramacy:2007:JSSOBK:v19i09} or the {\tt tgp} package
  vignette.}  Cokriging~\citep{verhoef:barry:1998},
co-regionalization~\citep{schmidt:gelfand:2003}, and other approaches
to modeling multivariate responses are obvious extensions, but lie
beyond the scope of the present work, and are not easily
parallelizable.  The MAP tree $\hat{\mathcal{T}}$, used for creating
sequential treed maximum entropy candidates, is taken from the treed
GP surrogates of each of the $m_Z$ responses in turn.

Chipman et al.~(1998) recommend running several parallel chains, and
sub-sampling from all chains in order better explore the posterior
distribution of the tree ($\mathcal{T}$).  Rather than run multiple
chains explicitly, the trial nature of adaptive sampling can be
exploited: at the beginning of each trial the tree can be randomly
pruned back.  Although the tree chain associated with an individual
trial may find itself stuck in a local mode of the posterior, in the
aggregate of all trials the chain(s) explore the posterior of
tree--space nicely.  Random pruning represents a compromise between
restarting and initializing the tree at a well--chosen starting place.
This {\em tree inertia} usually affords shorter burn--in of the MCMC at
the beginning of each trial.  The tree can also be initialized with a
run of the Bayesian treed LM, for a faster burn--in of the treed GP
chain.

Each trial executes at least $B$ burn--in and $T$ total MCMC sampling
rounds.  Samples are saved every $E$ rounds in order to reduce the
correlation between draws by thinning.  Samples of ALM and ALC
statistics need only be gathered every $E$ rounds, so thinning cuts
down on the computational burden as well.  If the {\em emcee} has no
responses waiting to be incorporated by BAS at the end of $T$ MCMC
rounds, then BAS can run more MCMC rounds, either continuing where it
left off, or after re-starting the tree (but saving all samples from
each chain).  New trials, with new candidates, start only when the
{\em emcee} is ready with a new finished response.  Such is the design
so that the computing time of each BAS trial does not affect the rate
of sampling.  Rather, a slow BAS runs fewer MCMC rounds per finished
response, and re-sorts candidates less often compared to a faster BAS.
A slower adaptive sampler yields less optimal sequential samples, but
still offers an improvement over na\"ive gridding.  In the experiments
that follow in the next two sections, the MCMC for the surrogate model
was run with $B=2000$, $T=7000$ and $E=2$.

\vspace{-0.5cm}
\section{Illustrative examples}
\label{sec:examples}
\vspace{-0.3cm}

In this section, sequential designs are built for three synthetic 
datasets with the treed GP LLM as a surrogate model.  The
supercomputer was simulated as having five independent nodes
which could provide responses for inputs in a time of 20 seconds
plus a random number of seconds having a Poisson distribution with
mean 20.  The examples in this section use the default prior specification
provided by the {\tt tgp} package, which involves using an improper
prior for $\bm{\beta}$ obtained by fixing $\bm{\beta}_0 = \mb{0}$ and
$\tau^2 = \infty$ in Eq.~(\ref{eq:model}) of Section \ref{sec:tgp}.  

\vspace{-0.1cm}
\subsection{1-d Synthetic Sinusoidal data}
\label{sec:as:sin}
\vspace{-0.1cm}

\begin{figure}[ht!]
\begin{center}
\vspace{-0.3cm}
\includegraphics[angle=-90,scale=0.28]{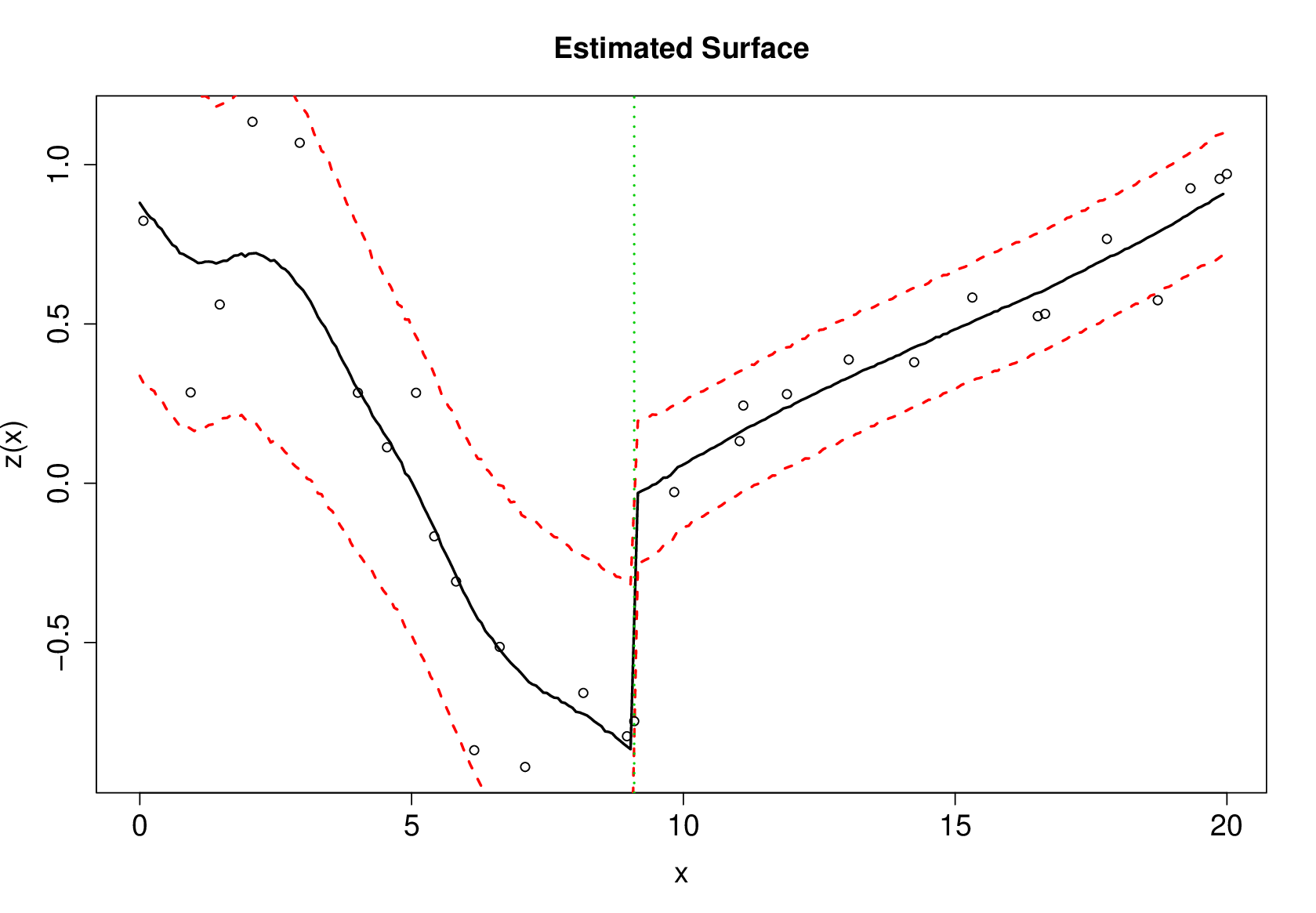}
\includegraphics[angle=-90,scale=0.28]{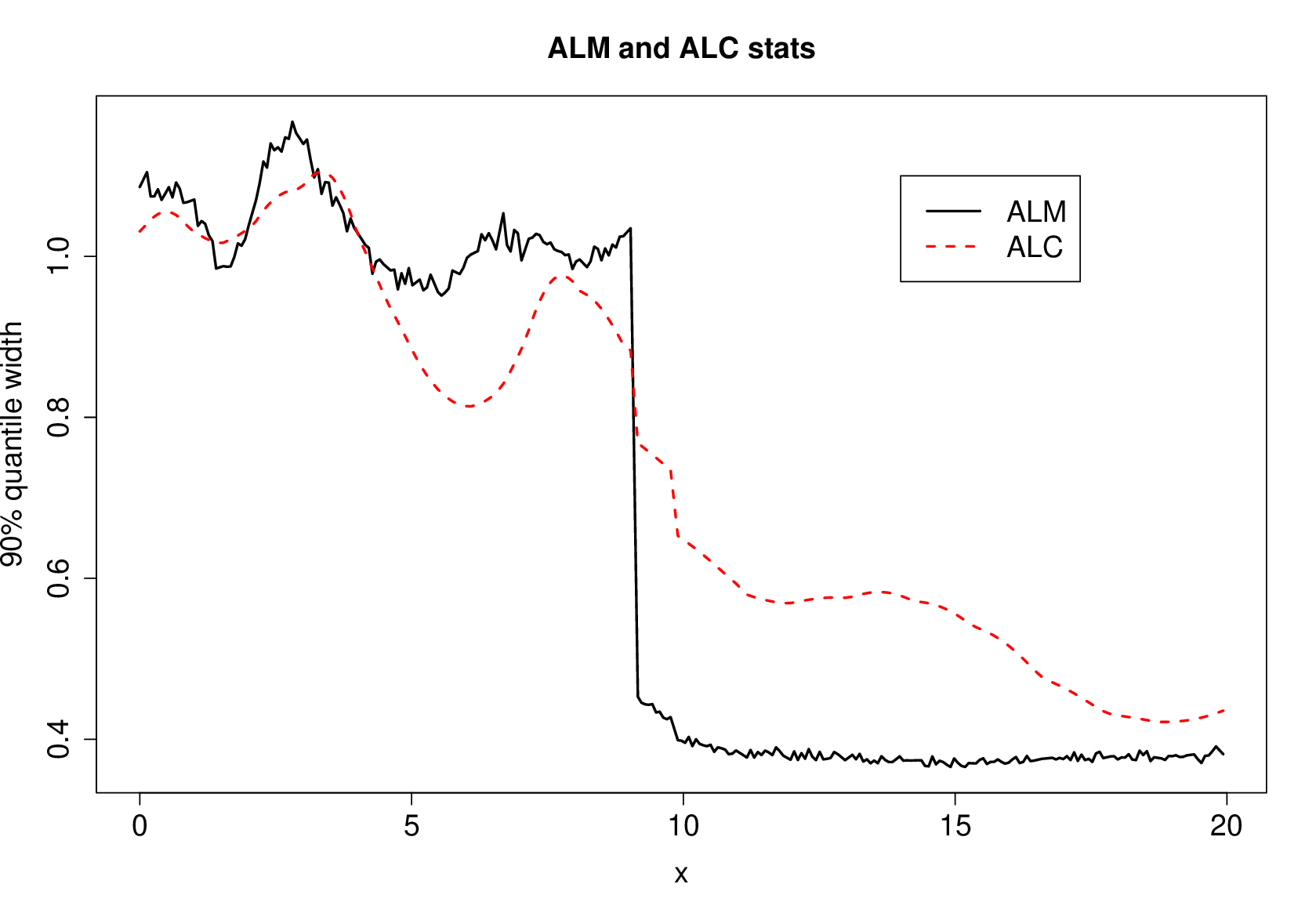}
\includegraphics[angle=-90,scale=0.28]{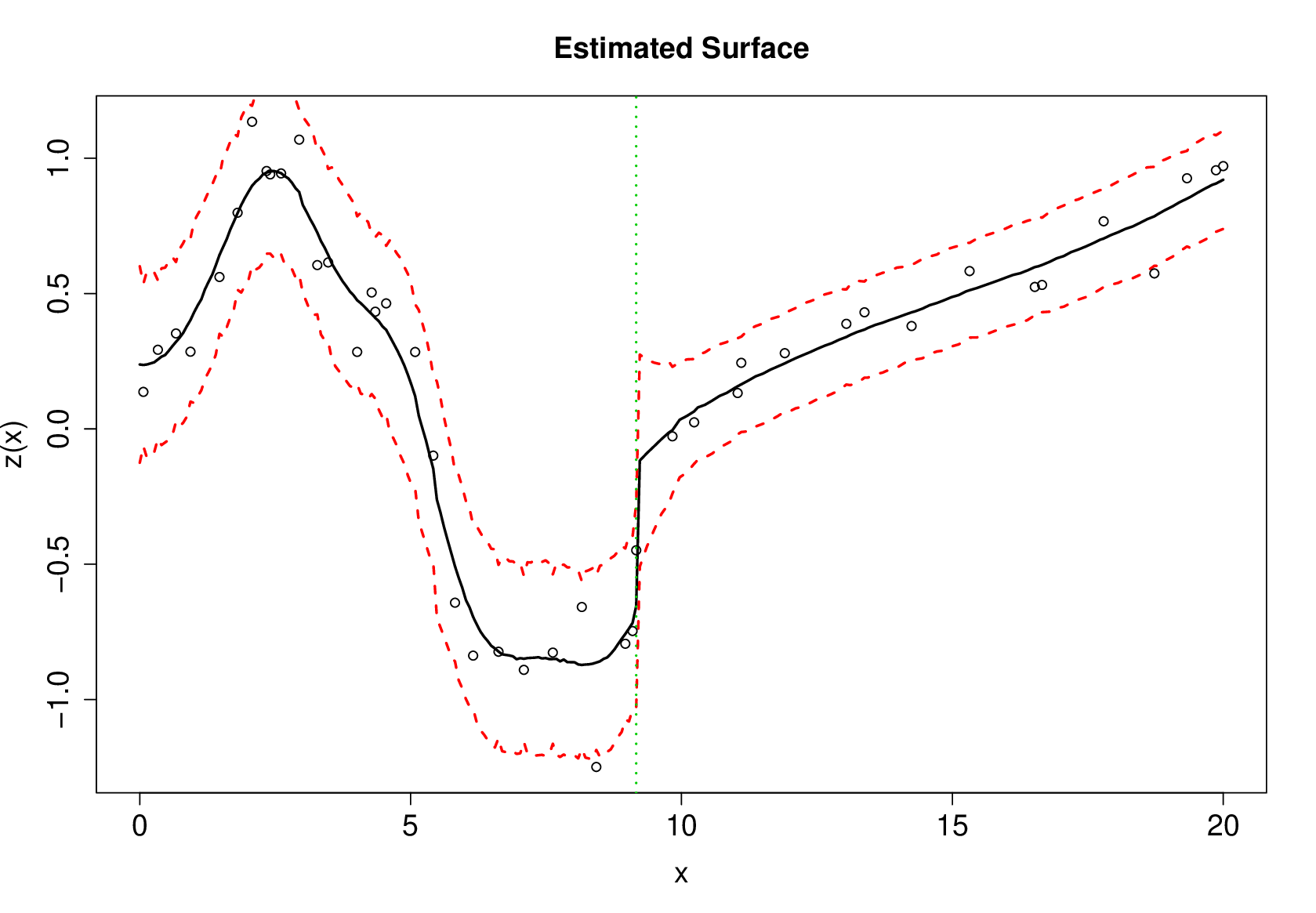}
\includegraphics[angle=-90,scale=0.28]{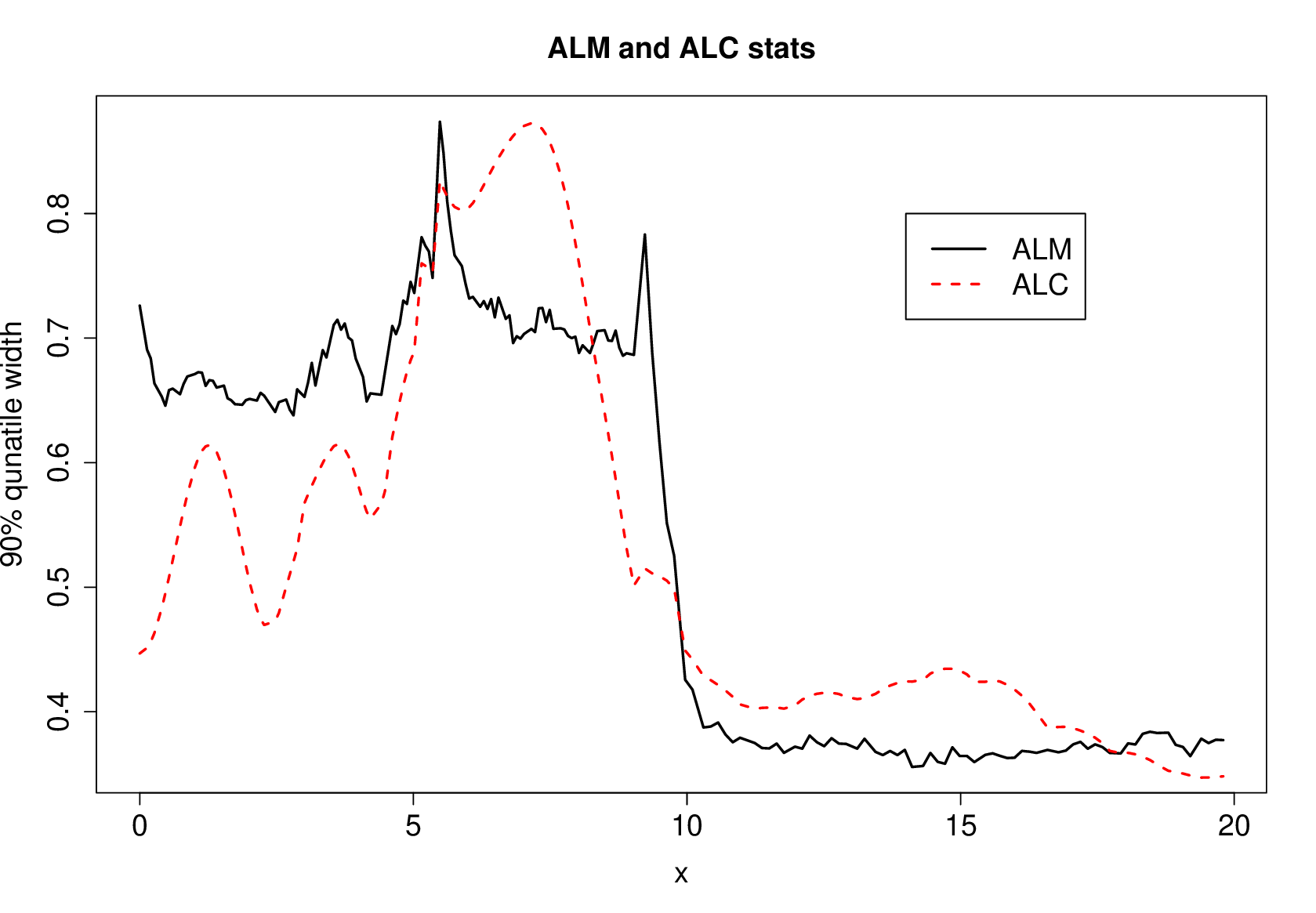}
\includegraphics[angle=-90,scale=0.28]{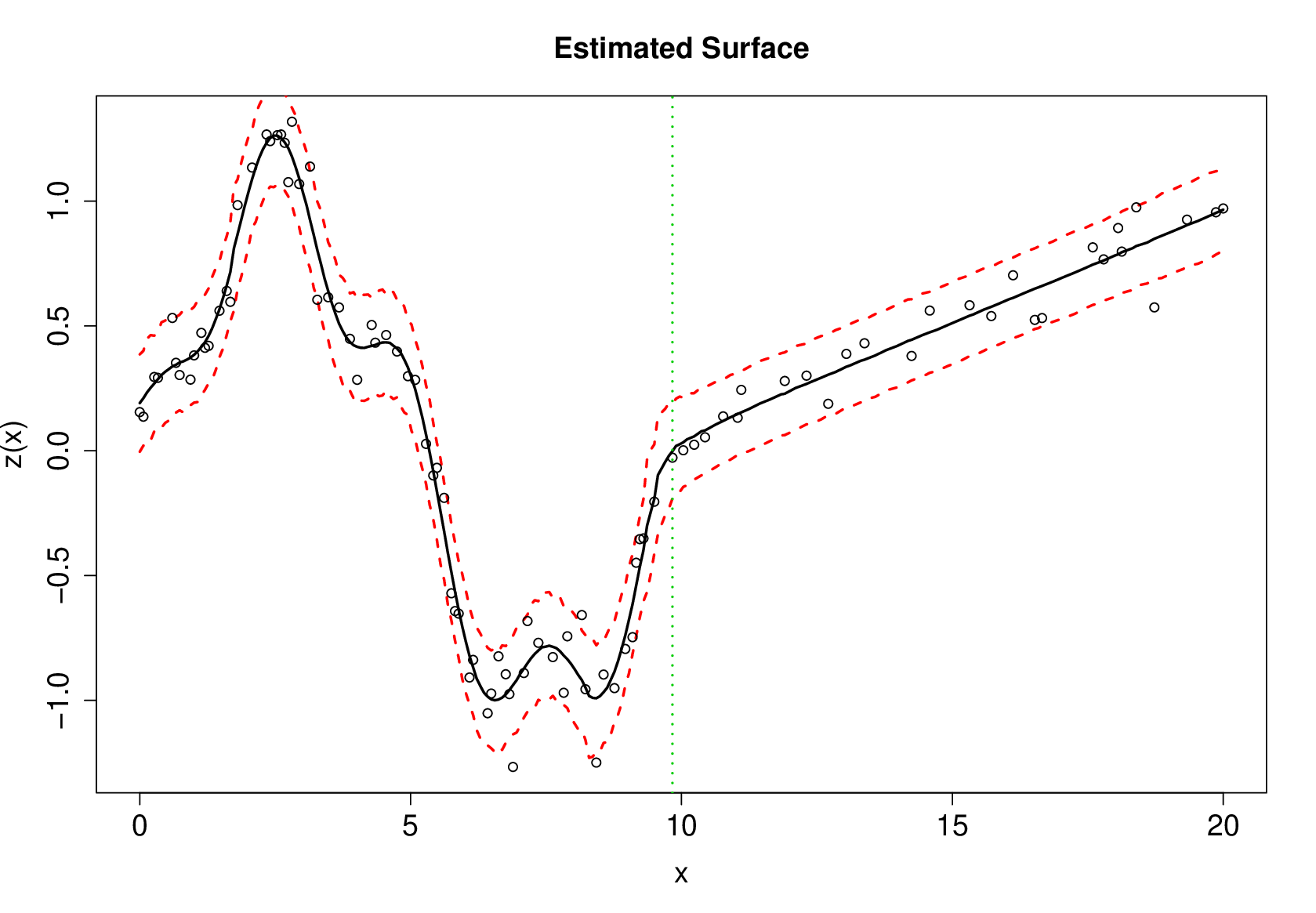}
\includegraphics[angle=-90,scale=0.28]{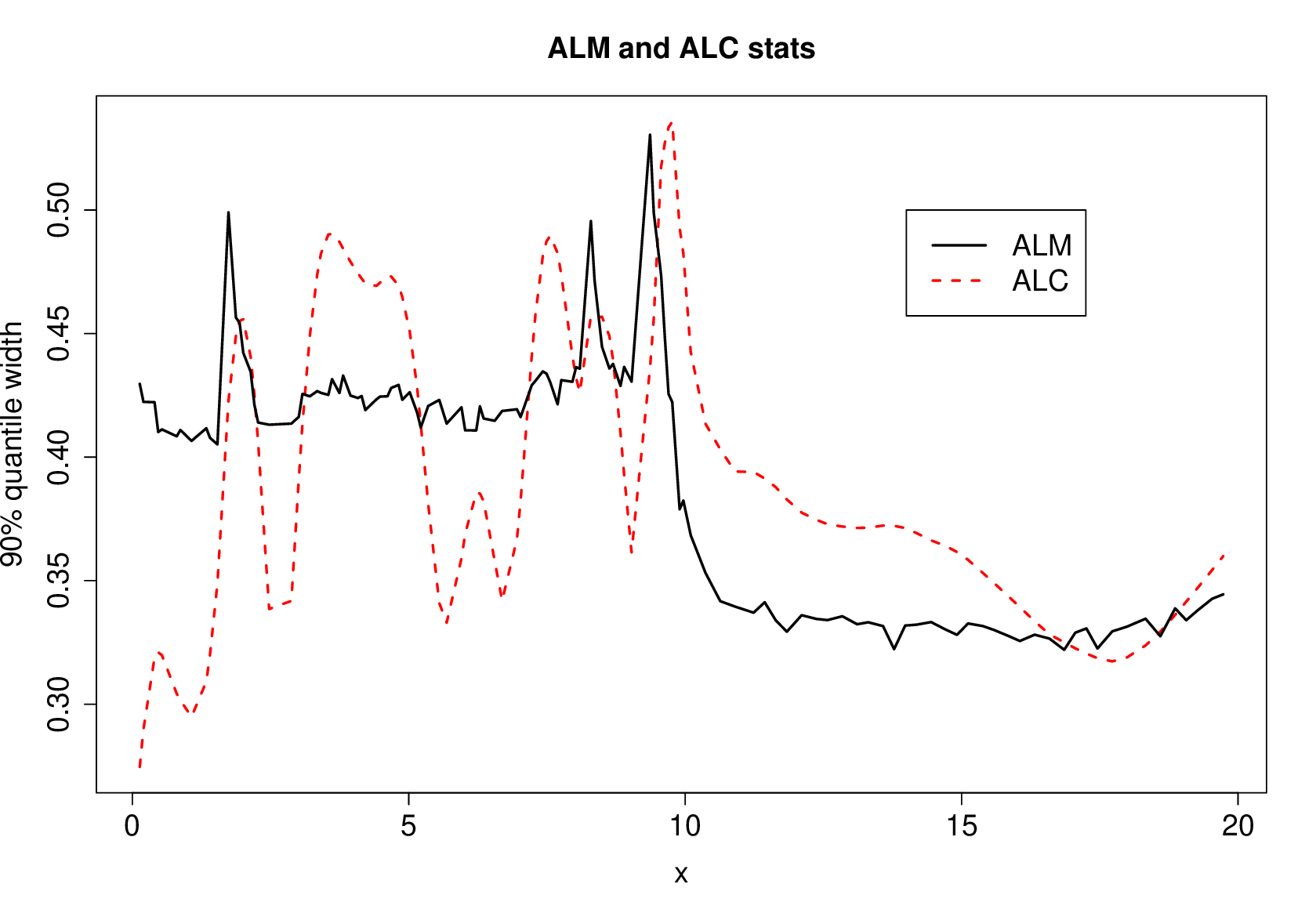}
\vspace{-0.2cm}
\caption[BAS on Sinusoidal data]{Sine data after 30 {\em (top)}, 45
  {\em (middle)}, and 97 {\em (bottom)} adaptively chosen samples.
  {\em Left:} posterior predictive mean and 90\% quantiles, and MAP
  partition $\hat{\mathcal{T}}$.  {\em Right:} ALM (black-solid) and
  ALC (red-dashed).}
\label{f:assin}
\end{center}
\vspace{-0.3cm}
\end{figure}

Consider some synthetic sinusoidal data first used by
\citet{higd:2002}, and then augmented by \citet{gra:lee:2008} to
contain a linear region: \vspace{-0.2cm}
\begin{equation}
    z(x) = \left\{ \begin{array}{cl}
    \sin\left(\frac{\pi x}{5}\right)
        + \frac{1}{5}\cos\left(\frac{4\pi x}{5}\right) & x < 10 \\
        x/10-1 & \mbox{otherwise},
   \end{array} \right. 
   \label{e:sindata}
\end{equation}
observed with $N(0,\sigma=0.1)$ noise.  Figure \ref{f:assin} shows
three snap shots, illustrating the evolution of BAS on this data using
the ALC algorithm with sequential treed maximum entropy candidates.
The first column shows the estimated surface in terms of posterior
predictive means (solid-black) and 90\% intervals (dashed--red).  The
MAP tree $\hat{\mathcal{T}}$ is shown as well.  The second column
summarizes the ALM and ALC statistics (scaled to show alongside ALM)
for comparison.  Ten samples from a sequential maximum entropy design
were used to start things off, and twenty candidates from a treed
maximum entropy design were proposed during each adaptive sampling
round.

The snapshot in the top row of Figure \ref{f:assin} was taken after
BAS had gathered a total of 30 samples, having learned that there is
probably one partition near $x=10$, with roughly the same number of
samples on each side.  Predictive uncertainty (under both ALM and ALC)
is higher on the left side than on the right.  ALM and ALC are in
relative agreement, however the transition of ALC over the partition
boundary is more smooth.  The ALM statistics are ``noisier'' than ALC
because the former is based on quantiles, and the latter on averages
(\ref{e:alcmean}).  Although both ALM and ALC are shown, only ALC was
used to select adaptive samples.  The middle row of Figure
\ref{f:assin} shows a snapshot taken after 45 samples were gathered,
where we can see that BAS has sampled more heavily in the sinusoidal
region (by a factor of two), and learned a great deal.  ALM and ALC
are in less agreement here than in the row above.  Also, ALC is far
less concerned with uncertainty near the partition boundary, than it
is, say, near $x=7$.  Finally, the snapshot in the bottom row of
Figure \ref{f:assin} was taken after 97 samples had been gathered.  By
now, BAS has learned about the secondary cosine structure in the {\em
  left--hand} region.  It has focused almost three times more of its
sampling effort there.  ALM and ALC agree that there is high
uncertainty near the partition boundary $(\hat{\mathcal{T}})$, but
otherwise disagree about where to sample next.  
Any further sampling would yield only marginal improvements since this
final surface, in the {\em bottom--left} panel, is a very good
approximation to the truth.

In summary, the left panels of Figure \ref{f:assin} track the treed GP
model's improvements in its ability to predict the mean, via the
increase in resolution from one figure to the next.  From the scale of
$y$-axes in the right column one can see that as more samples are
gathered, the variance in the posterior predictive distribution of the
treed GP decreases as well.  Despite the disagreements between ALM and
ALC on individual iterations during the evolution of BAS, it is
interesting to note that difference between using ALC and ALM on this
dataset is negligible.  This is likely due to the high quality of the
candidates $\tilde{\mb{X}}$, from a sequential treed maximum entropy
design, which prevent the clumping behavior that tends to hurt ALM, but
to which ALC is somewhat less prone.

Perhaps the best illustration of how BAS learns and adapts over time
is to compare it to something that is, ostensibly, less adaptive.
\citet{glm:04} show how the mean-squared error (MSE) of BAS evolves
over time on a similar dataset, but in a serial setting where only one
sample is taken at a time, and the surrogate model is allowed to
re-fit before the next (single) adaptive sample is chosen.  They show
how the MSE of BAS decreases steadily as samples are added, despite
that fewer points are added in the linear region, yielding a
sequential design strategy which is two times more efficient than LHS.
They also show how BAS measures up against ALM and ALC, as implemented
by Seo et al.~(2000)---with a stationary GP surrogate model.  Seo et
al.~make the very powerful assumption that the correct covariance
structure is known at the start of sampling.  Thus, the model need not
be updated in light of new responses.  Alternatively, BAS quickly
gathers enough samples to {\em learn} the partitioned covariance
structure, after which it outperforms ALM and ALC based on a
stationary model.

\vspace{-0.1cm}
\subsection{2-d Synthetic Exponential data}
\label{sec:as:exp} 
\vspace{-0.1cm}

\begin{figure}[ht!]
\begin{center}
\includegraphics[scale=0.35,trim=40 190 0 100,angle=-90]{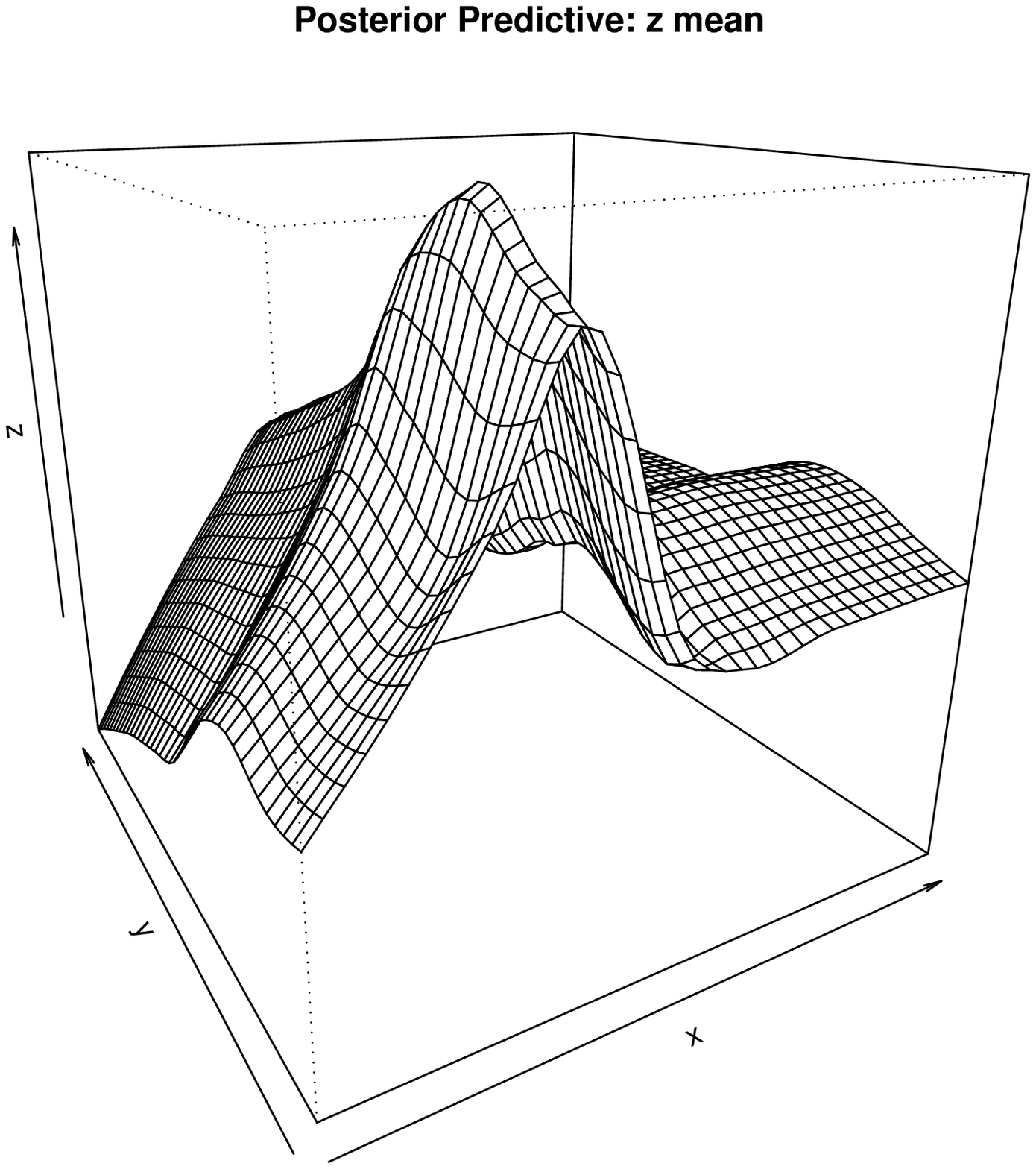}
\includegraphics[scale=0.32,trim=20 20 30 20,angle=-90]{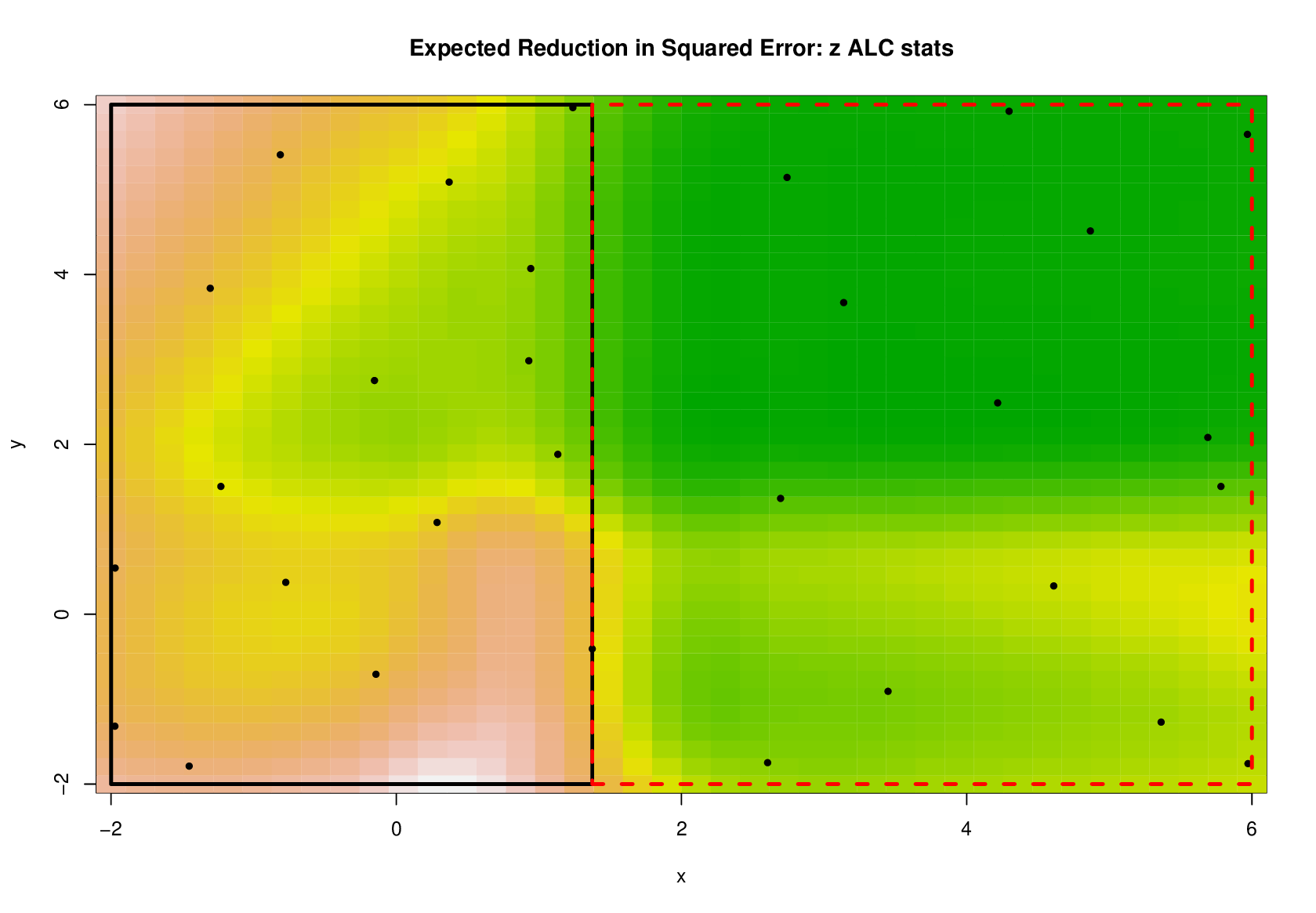} \\
\includegraphics[scale=0.35,trim=40 190 0 100, angle=-90]{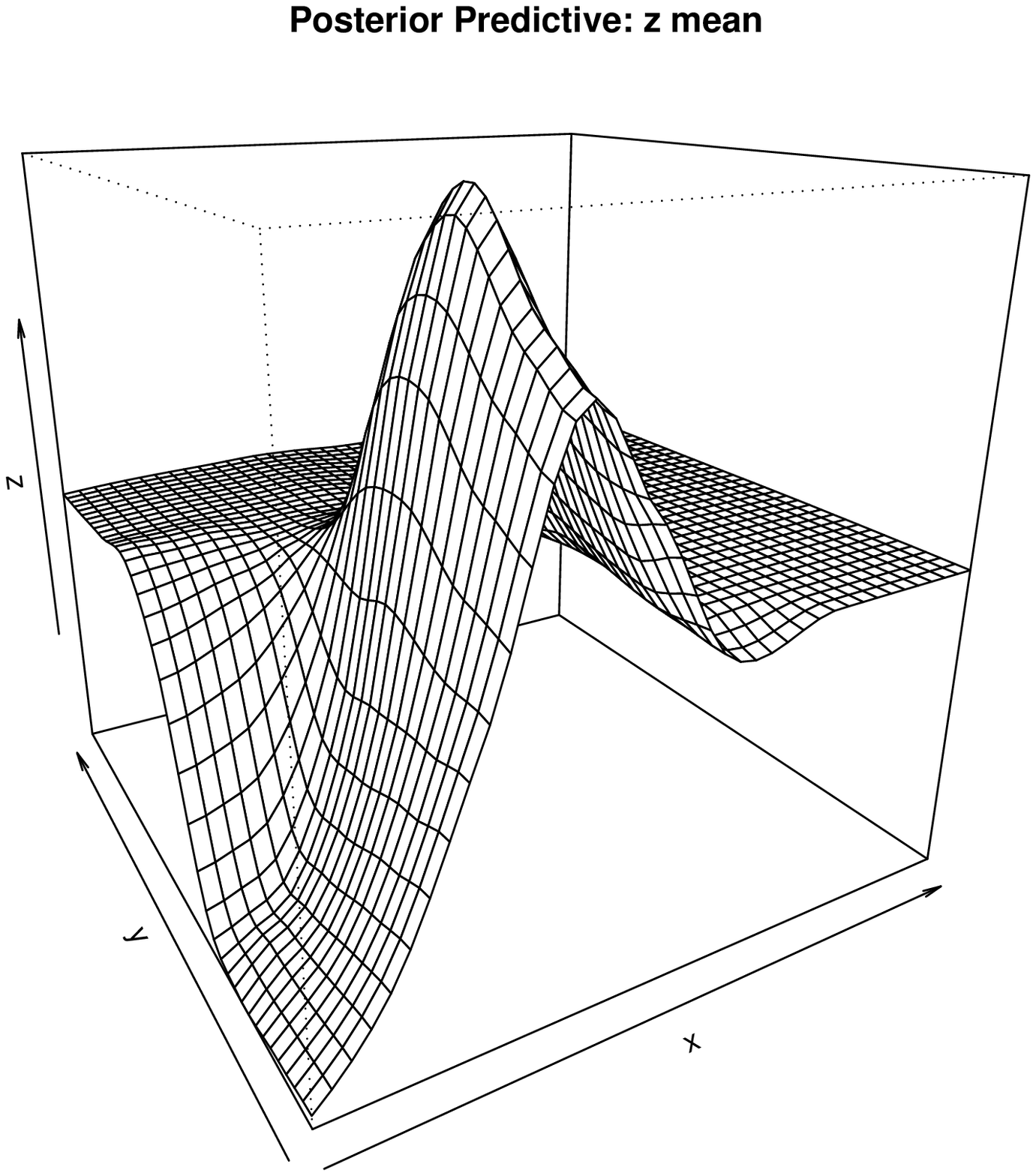}
\includegraphics[scale=0.32,trim=25 20 30 20,angle=-90]{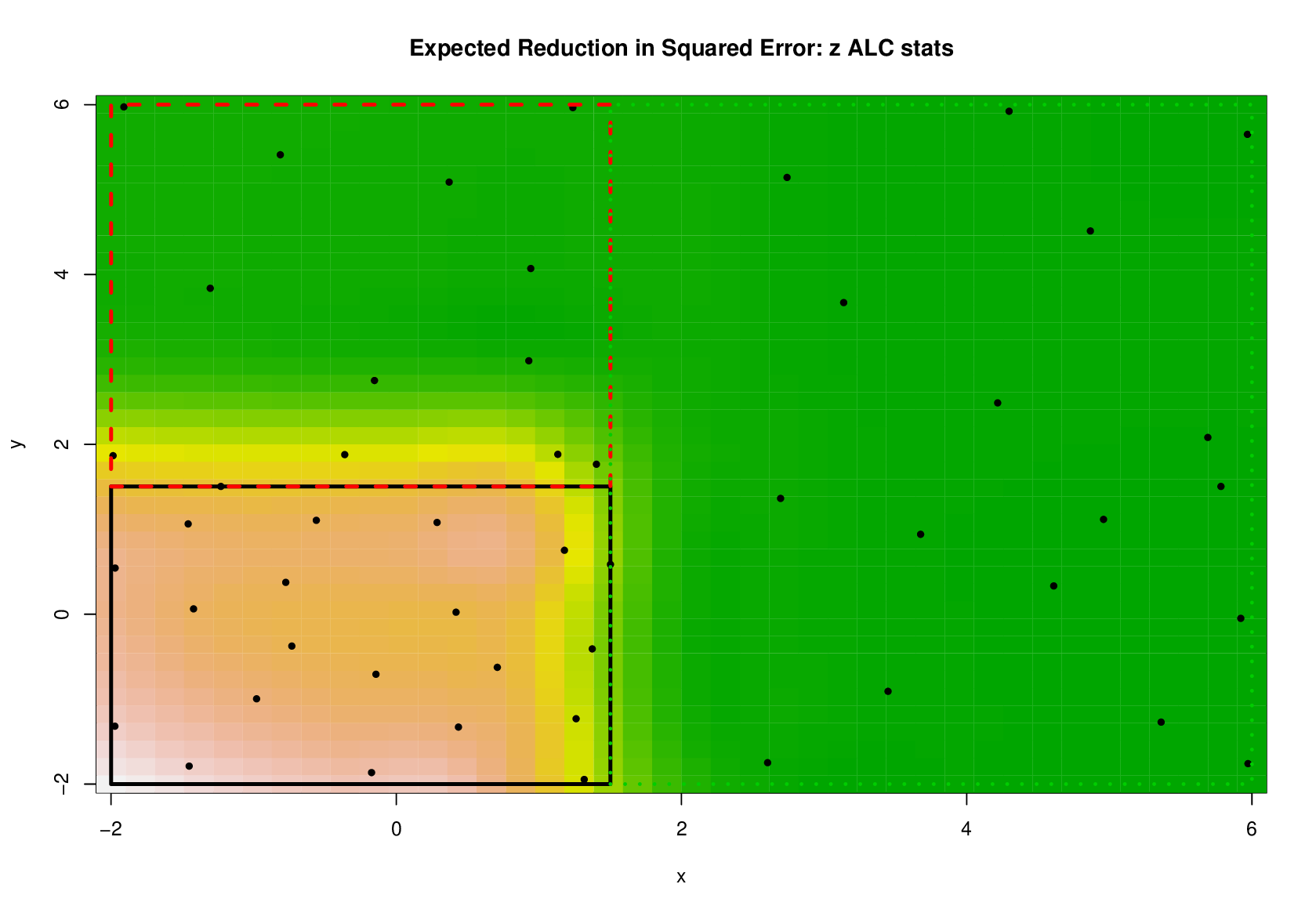}\\
\includegraphics[scale=0.35,trim=40 190 0 100,angle=-90]{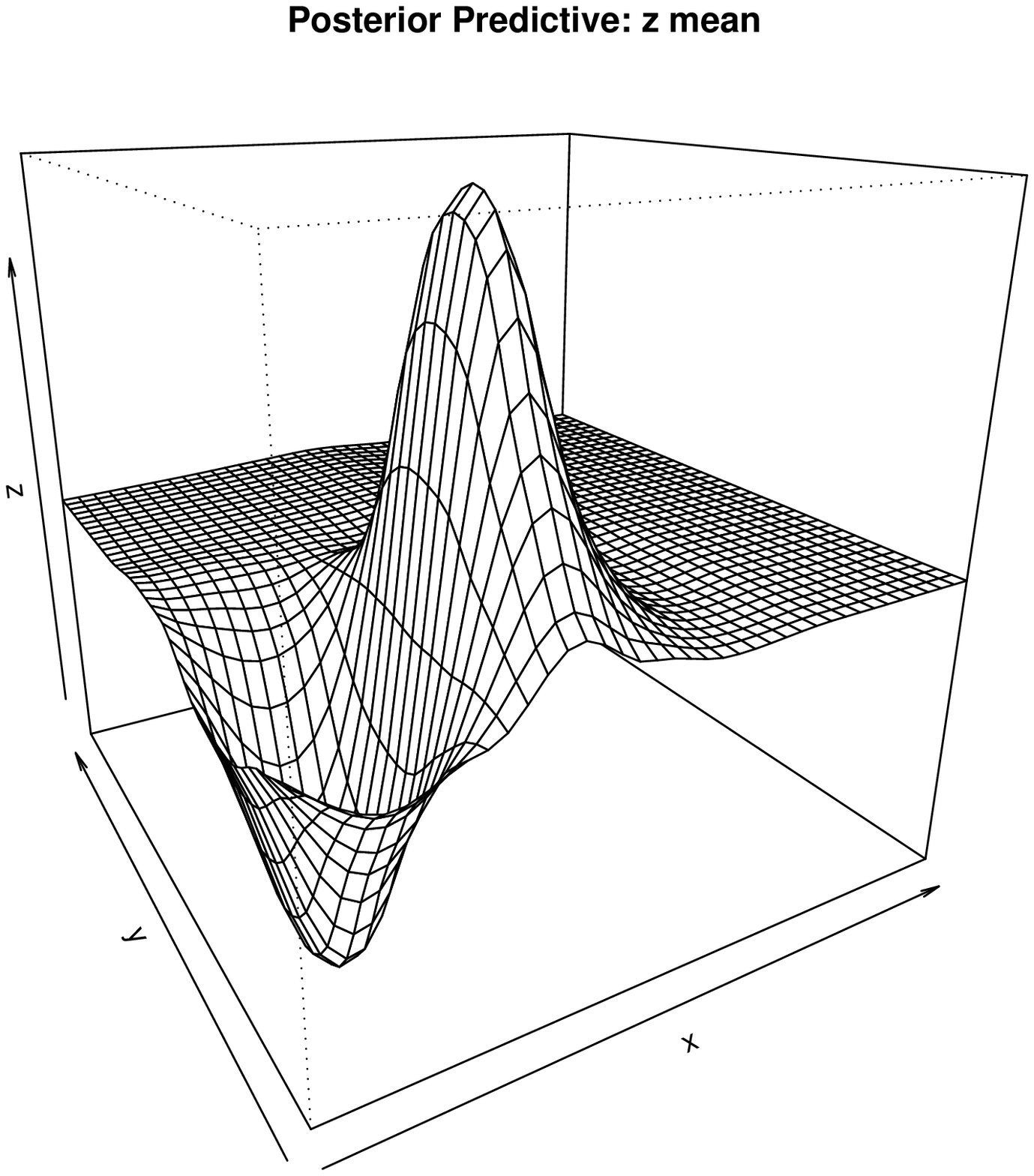}
\includegraphics[scale=0.32,trim=25 20 30 20,angle=-90]{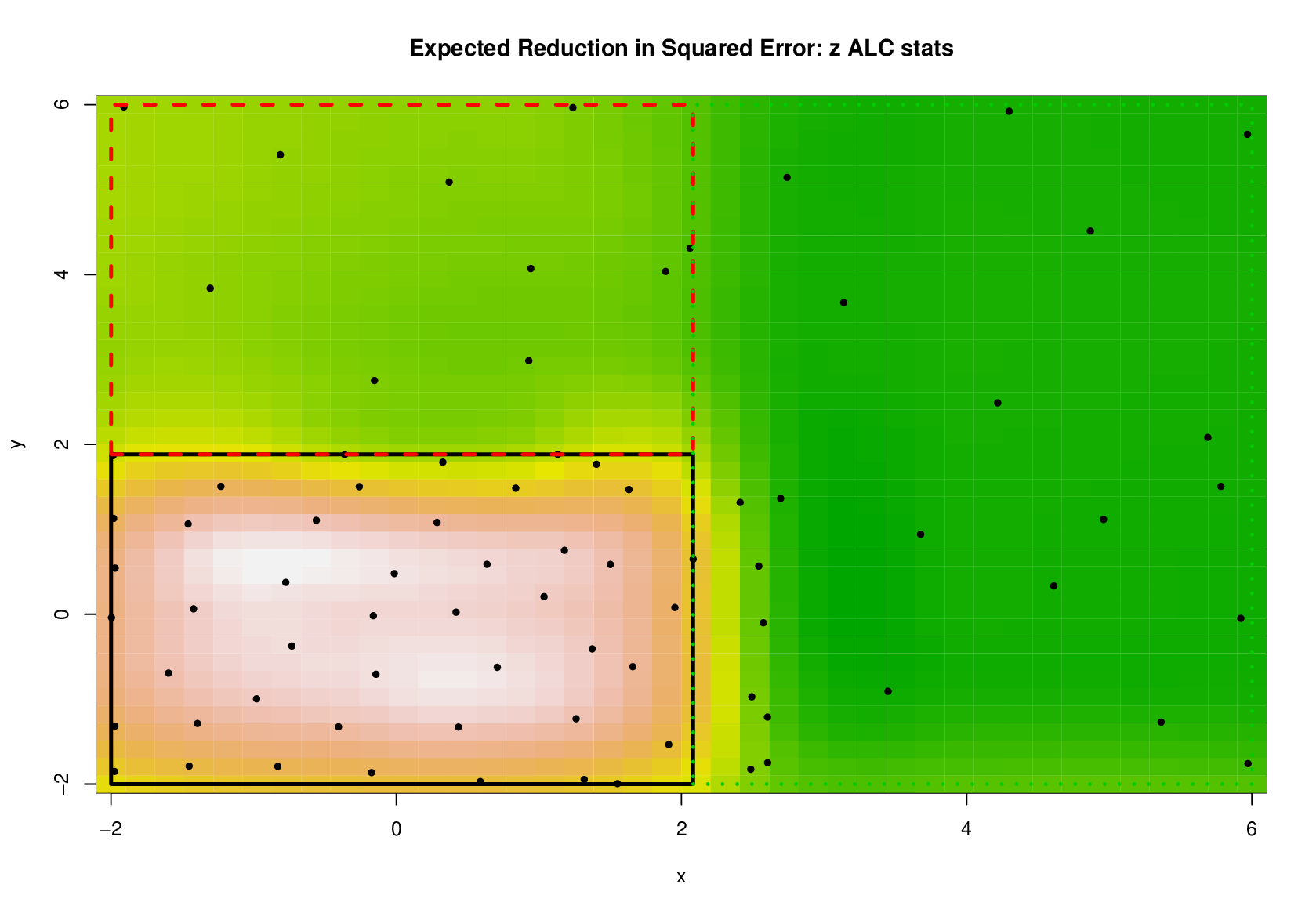}\\
\vspace{-0.75cm}
\end{center}
\caption[BAS on Exponential data]{ Exponential data after 30 {\em (top)},
50 {\em (middle)} and 80 {\em (bottom)} adaptively chosen samples.  
{\em Left:} posterior predictive mean surface; 
{\em Right:}  ALC criterion surface, with MAP
 tree $\hat{\mathcal{T}}$ and samples $X$ overlayed.
\label{f:asexp}}
\vspace{-0.2cm} 
\end{figure}

The nonstationary treed GP surrogate model has an even greater impact
on adaptive sampling in a higher dimensional input space.  For an
illustration, consider the domain $[-2,6] \times [-2,6]$ wherein the
true response is given by $ z(\mb{x}) = x_1 \exp(-x_1^2 - x_2^2)$,
observed with $N(0,\sigma=0.001)$ noise.  We take an initial set of 16
configurations from a maximum entropy design, and twenty new
candidates (from a sequential treed maximum entropy design) are used
in each adaptive sampling round.  The top row of Figure \ref{f:asexp}
shows a snapshot after 30 adaptive samples have been gathered with BAS
under the ALC algorithm.  Room for improvement is evident in the mean
predictive surface ({\em left} column).
The second column shows the ALC surface, and the single partition of
$\hat{\mathcal{T}}$, with samples evenly split between the two
regions.  Observe in the ALC plot that the model also considers a tree
with a single split along the other axis, indicating good mixing of
the reversible jump Markov chain in tree space.

After 50 adaptive samples have been selected ({\em second--row} of
Figure \ref{f:asexp}), the situation is greatly improved.  ALC asserts
that the first quadrant is most interesting, and as a result
(adaptive) sampling is higher in this region.  The dots in Figure
\ref{f:treeddopt} illustrate the candidates from a sequential treed
maximum entropy design used during this round.  [Notice that the
$\mb{X}$ locations (circles) in Figure \ref{f:treeddopt} match the
dots in the center row of Figure \ref{f:asexp}.]  Finally, the bottom
row of Figure \ref{f:asexp} shows the snapshot taken after 80 adaptive
samples.
More than 54\% of the samples are located in the
first quadrant which occupies only 25\% of the total input space.  
As before, the final surface shown in the {\em bottom--left} of the
figure is a very good approximation to the truth.

When comparing to LHS, ALM, and ALC with stationary GPs, much the same
can be said here as with the sinusoidal data.  \citet{glm:04} show
that the MSE of BAS decreases steadily as samples are added in a
serial fashion, despite that most of the sampling occurs in the first
quadrant, and that it is at least two--times more efficient than LHS.
Crucially, the exponential data are not defined by step functions, in
contrast with the sinusoidal data.  Transitions between partitions are
smooth.  Thus it takes BAS longer to learn about $\mathcal{T}$---which
in this case can be thought of as a design tool rather than a model
assumption (since the data are well fit by a stationary GP)---and the
corresponding three GP models in each region of $\hat{\mathcal{T}}$.
Once it does however (after about 50 samples) BAS outperforms the (in
hindsight) well--parameterized stationary model with ALM.

\begin{table}[ht]
\begin{center}
\begin{tabular}{lllr}
  \hline
model & cands & as & rmse \\
  \hline
btgp & lh & alc & 0.00346 \\
btgp & lh & alm & 0.00365 \\
btgpllm & lh & alm & 0.00366 \\
bcart & tme & alc & 0.00430 \\
btgpllm & lh & alc & 0.00459 \\
bcart & tme & alm & 0.00531 \\
btgp & tme & alm & 0.00561 \\
btgpllm & tme & alm & 0.00678 \\
btgpllm & tme & alc & 0.00722 \\
btgp & tme & alc & 0.00765 \\
btlm & tme & alm & 0.00874 \\
bcart & lh & alm & 0.00903 \\
bcart & lh & alc & 0.00934 \\
btlm & lh & alm & 0.00989 \\
\hline
\end{tabular}
\hspace{0.5cm}
$\cdots$
\hspace{0.5cm}
\begin{tabular}{lllr}
  \hline
model & cands & as & rmse \\
  \hline
btlm & lh & alc & 0.01148 \\
btlm & tme & alc & 0.01572 \\
bgp & lh & alm & 0.03519 \\
btgp & me & alc & 0.03747 \\
bcart & me & alm & 0.03885 \\
bcart & me & alc & 0.04002 \\
btgpllm & me & alc & 0.04028 \\
btgp & me & alm & 0.04122 \\
btgpllm & me & alm & 0.04245 \\
btlm & me & alc & 0.04269 \\
btlm & me & alm & 0.04344 \\
bgp & lh & alc & 0.04929 \\
bgp & me & alm & 0.05090 \\
bgp & me & alc & 0.05553 \\
   \hline
\end{tabular}
\end{center}
\caption{
  Comparing a combination of five models: Bayesian CART, treed
  linear models, (stationary) GP, treed GP, and treed GP LLM (labeled bcart,
  btlm, bgp, btgp, btgpllm); three ways of generating AS
  candidates: LHS, sequential maximum entropy and sequential treed 
  maximum entropy
  (lh, me, tme); and two AS heuristics ALC and ALM, in terms
  of RMSE to the truth.
  Observe that the bgp/tme combination is not run since the
  stationary GP model does not provide a MAP tree as is required 
  for sequential treed maximum entropy design.  Therefore there are
  28 combinations instead of 30.
  \label{f:expres}}
\end{table}

To highlight the benefits of using a treed model in sequential design,
we consider a deeper comparison on this data with results summarized
in Table \ref{f:expres}.  The comparison involves combinations of five
models: Bayesian CART, treed linear models, (stationary) GP, treed GP,
and treed GP LLM; three ways of generating AS candidates: LHS, maximum
entropy and treed maximum entropy; and two adaptive sampling
heuristics: ALC and ALM.  The table shows RMSE to the truth as
evaluated on a dense grid, for 30 repeated BAS runs each starting with
a random initial maximum entropy design of 20 configurations, and then
55 samples chosen adaptively (for 75 total).  For fairness, the final
RMSE calculation (in each case) is based on the predictive means
sampled from a full treed GP LLM model on a dense grid of predictive
locations, regardless of the method used for sequential design.  The
table is sorted on the fourth column (RMSE).  Somewhat surprisingly,
Bayesian CART does really well if its candidates come from a
sequential treed maximum entropy design.\footnote{Note that the using
  the full treed GP LLM for the RMSE calculation is crucial here.  Had
  Bayesian CART been used instead it would have been ranked much
  lower.}  Also, ALM and ALC perform about equally as well as one
another on this data, though we suspect that ALC would do better than
ALM if the data were heteroskedastic, in which case ALM would
concentrate samples in the high noise region even if the mean in that
region is tame.  Finally, LHS candidates do better than ones from a
sequential maximum entropy design (with the notable exception of
Bayesian CART).  The non-treed (stationary) GP model and non-treed
maximum entropy designs are the worst in the study.  That the treed GP
does better than the treed GP LLM is perhaps to be expected as there
is nothing at all linear about this dataset.

\vspace{-0.1cm}
\subsection{Six-dimensional example}
\label{sec:as:sixd} 
\vspace{-0.1cm}

As an example of a higher-dimensional problem, we present a
6-d example, with true response \vspace{-0.4cm}
\begin{equation} 
 z(x_1, x_2, x_3, x_4, x_5, x_6) = \exp\left\{\sin\left(\left[0.9*(x_1
 + 0.48)\right]^{10}\right)\right\} + x_2x_3 + x_4 \,.
\label{eq:sixd}
\vspace{-0.2cm}
\end{equation}
This function has four active variables.  It is of continuously
varying wiggliness in the first dimension where the smoothness varies
over the space without any natural threshold.  The treed GP will
usually partition on this dimension, typically somewhere between 0.6
and 0.85, which will allow more of the adaptive sampling effort to be
put on the more quickly oscillating part near $x_1 = 1$.  The response
is smooth but non-linear in the second and third dimensions, and
linear in the fourth dimension.  The final two variables are pure
noise, which the treed GP will need to learn about adaptively.

\begin{table}[ht!]
\centering
\begin{tabular}{lllr}
  \hline
method & Avg(rmse) & SE(rmse) \\
  \hline
btgpllm-linburn/tme/alc & 0.02871943 & 0.0006446596 \\
btgpllm/tme/alc & 0.03217273 & 0.0037997994 \\
no adaptive sampling & 0.03598135 & 0.0034438090
\end{tabular}
\caption{RMSE to the truth as evaluated on random LHSs of size 1000 in $[0,1]^6$, summarized for 10 repeated AS runs 
  on the 6-d example. }
\label{t:sixd}
\end{table}

Our experiment allows the inputs to vary in $[0,1]^6$ and the response
in (\ref{eq:sixd}) is observed with $N(0,\sigma=0.05)$ noise.  We used
a similar artificial clustered simulation environment to the one
described in Section \ref{sec:as:methods}.  At any time there are five
nodes available to evaluate responses, which finish in no sooner than
3 minutes, plus a random number of seconds distributed as Pois$(180)$.
Table \ref{t:sixd} shows average RMSE to the truth as evaluated on 10
random LHSs of size 1000 in $[0,1]^6$ (with standard errors), for 10
repeated BAS runs.  Each run starts with a random initial set of 400
configurations from a maximum entropy design, and then 400 samples are
chosen adaptively.  We compare the Bayesian treed GP LLM model with
ALC, both with and without LM burn--in, to a non-adaptive maximum
entropy design of size 800.  A stationary (non-treed) GP model was
excluded from the comparison due to time constraints.  As before, we
use the full treed GP LLM model to calculate RMSEs after the adaptive
sampling run(s), for fairness.  Though there is some variation across
the 10 runs, the RMSE values obtained in each run were always
row--ordered as they are, in the table, with the non-adaptive method
coming last, and the treed GP LLM version with linear burn--in and ALC
coming first.  That is, although the differences between the average
RMSEs in the table appear to be modest, the improvement obtained by
BAS is statistically significant.

\vspace{-0.3cm}
\section{LGBB CFD experiment}
\label{sec:lgbb}
\vspace{-0.3cm}

The final experiment is our motivating example, a computational fluid
dynamics simulator of a proposed reusable NASA launch vehicle, called
the Langley Glide--Back Booster (LGBB).  Three input parameters are
varied (side slip angle, speed, and angle of attack), and for each
setting of the input parameters, six outputs (lift, drag, pitch,
side-force, yaw, and roll) are monitored.  All six responses are
computed simultaneously.  In a previous experiment, a supercomputer
interface was used to launch runs at over 3,250 input configurations
in several hand--crafted batches.  Figure \ref{f:cfdinit1} plots the
resulting lift response as a function of Mach (speed) and alpha (angle
of attack), with beta (side-slip angle) fixed to zero.  A more
detailed description of this system and its results are provided by
\citet{rog03}.  Some preliminary adaptive sampling of this data
appeared in \citet{glm:04}, although that paper dealt with only a
single output, considered only one-at-a-time updates, involved only a
simulation of a computer experiment, and only resampled points from
the hand--crafted initial run of 3,250.  Here we describe the
development of a live sequential design in the fully asynchronous NASA
environment.  In a separate, non-adaptive, analysis of this data
\citet{gra:lee:2008} noticed that the noise structure was
heteroskedastic, so we have chosen to use the ALC statistic to guide
the adaptive sampling.

BAS for the LGBB is illustrated pictorially by the remaining figures
in this section.  The experiment was implemented on the NASA
supercomputer {\tt Columbia}---a fast and highly parallelized
architecture, but with an extremely variable workload.  The {\em
  emcee} algorithm of Section \ref{sec:as:comp} was designed to
interface with {\tt AeroDB}, a database queuing system used by NASA to
submit jobs to {\tt Columbia}, and a set of CFD simulation codes
called {\tt cart3d}.  To minimize impact on the queue, the {\em emcee}
was restricted to ten submitted simulation jobs at a time.  Candidate
locations were sub-sampled from a 3-d grid consisting of 37,909
configurations.  The design was initialized with 30 candidates from a
maximum entropy design, and 100 new candidates (from a treed maximum
entropy design) were proposed during each AS round.  We use full
hierarchical prior for $\bm{\beta}$, described by Eq.~(\ref{eq:model})
in Section \ref{sec:tgp}, by augmenting the default with the augment
\verb!bprior = "b0"!.  The pooling of means implied by the
hierarchical prior is appropriate for this data since it is believed
that the vast expanse of the response surface---for speeds greater
than Mach 1---is largely homogeneous.

\begin{figure}[ht!]
\begin{center}
\begin{tabular}{cc}
\includegraphics[angle=-90,scale=0.28]{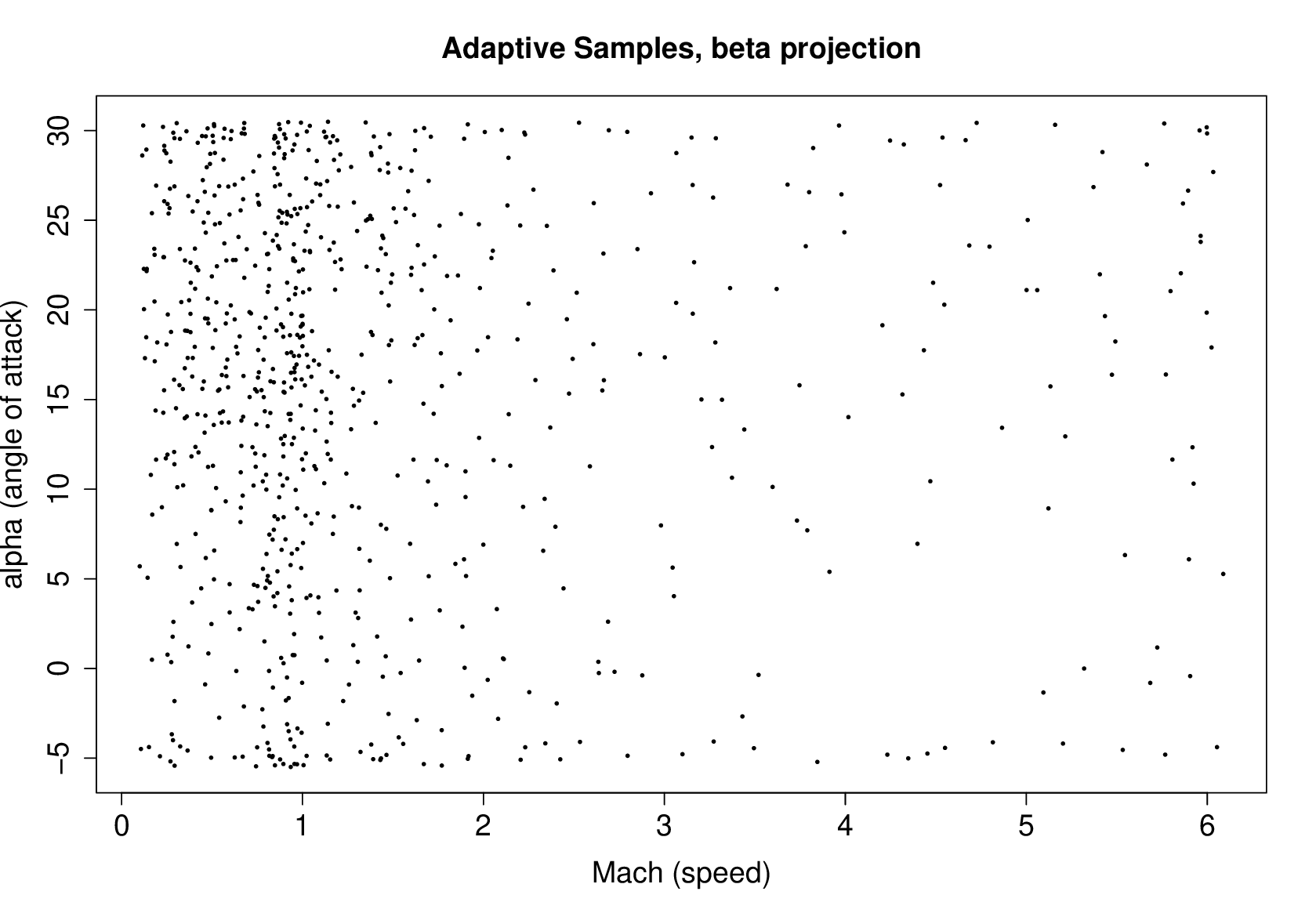} &
\includegraphics[scale=0.28,angle=-90]{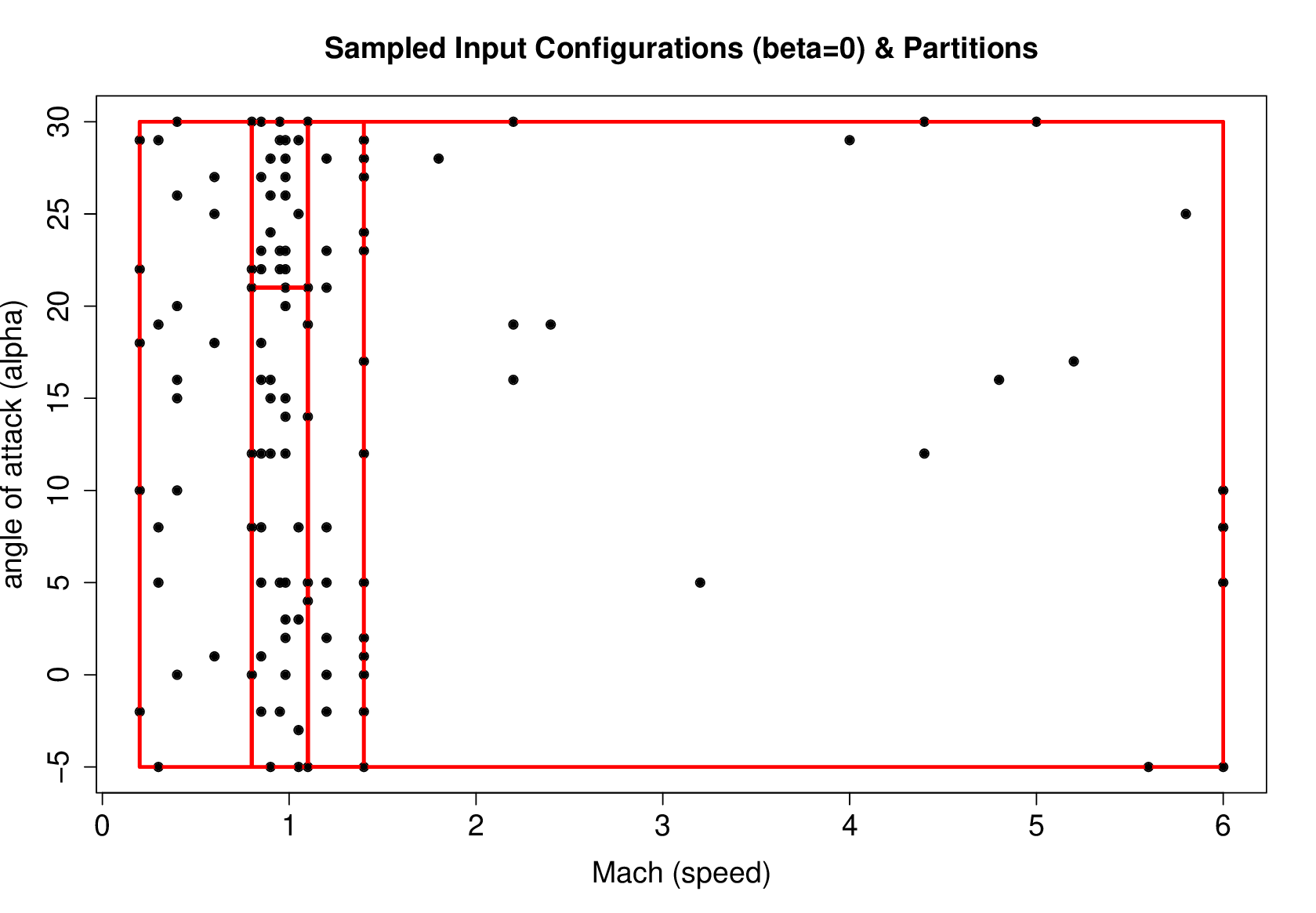}\\
\includegraphics[angle=-90,scale=0.28]{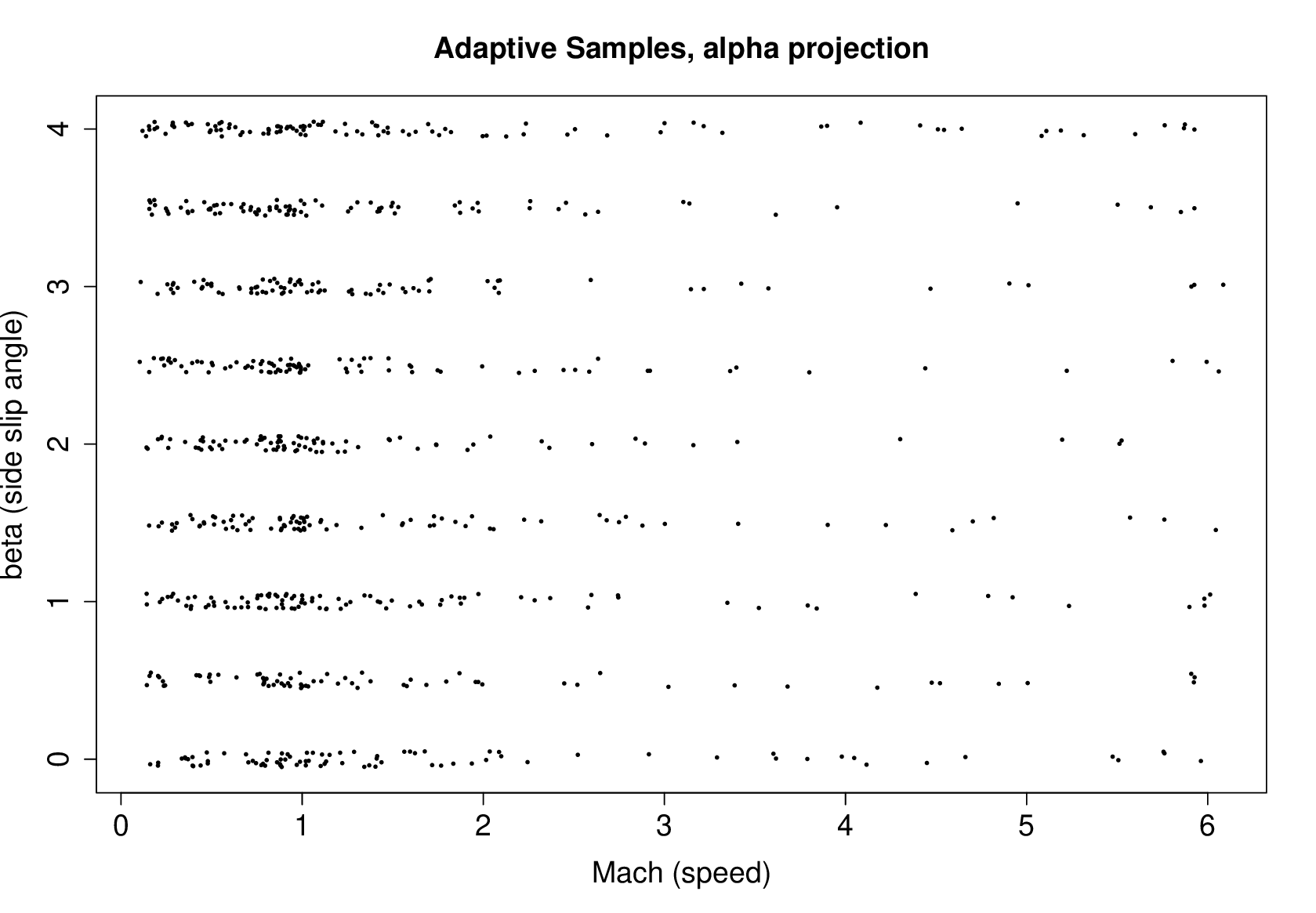} 
\end{tabular}
\vspace{-0.1cm}
\caption[LGBB full adaptively sampled configurations]{ Adaptively
  sampled configurations projected over beta (side-slip angle; {\em
    top--left}), for fixed beta = 0 ({\em top--right}) with MAP partition
  $\hat{\mathcal{T}}$, and then over alpha (angle of attack; {\em
    bottom--left}).}
\label{f:lgbbpoints}
\end{center}
\vspace{-0.3cm}
\end{figure}

Figure \ref{f:lgbbpoints} shows the 780 configurations sampled by BAS
for the LGBB experiment.  The left panel shows locations as a function
of Mach (speed) and alpha (angle of attack), projecting over beta
(side slip angle); the right panel shows Mach versus beta, projecting
over alpha; the middle panel shows the beta~=~0 slice.  NASA
recommended treating beta as discrete, so we used a set of values
which they provided.  We can see that most of the configurations
chosen by BAS were located near Mach one, with highest density for
large alpha.  Samples are scarce for Mach greater than two and are
relatively uniform across all beta settings.  A small amount of random
noise has been added to the beta values in the plots ({\em
  bottom--left}) for visibility purposes.

\begin{figure}[ht!]
\begin{center}
\vspace{-0.2cm}
\includegraphics[scale=0.30,trim=100 0 0 0]{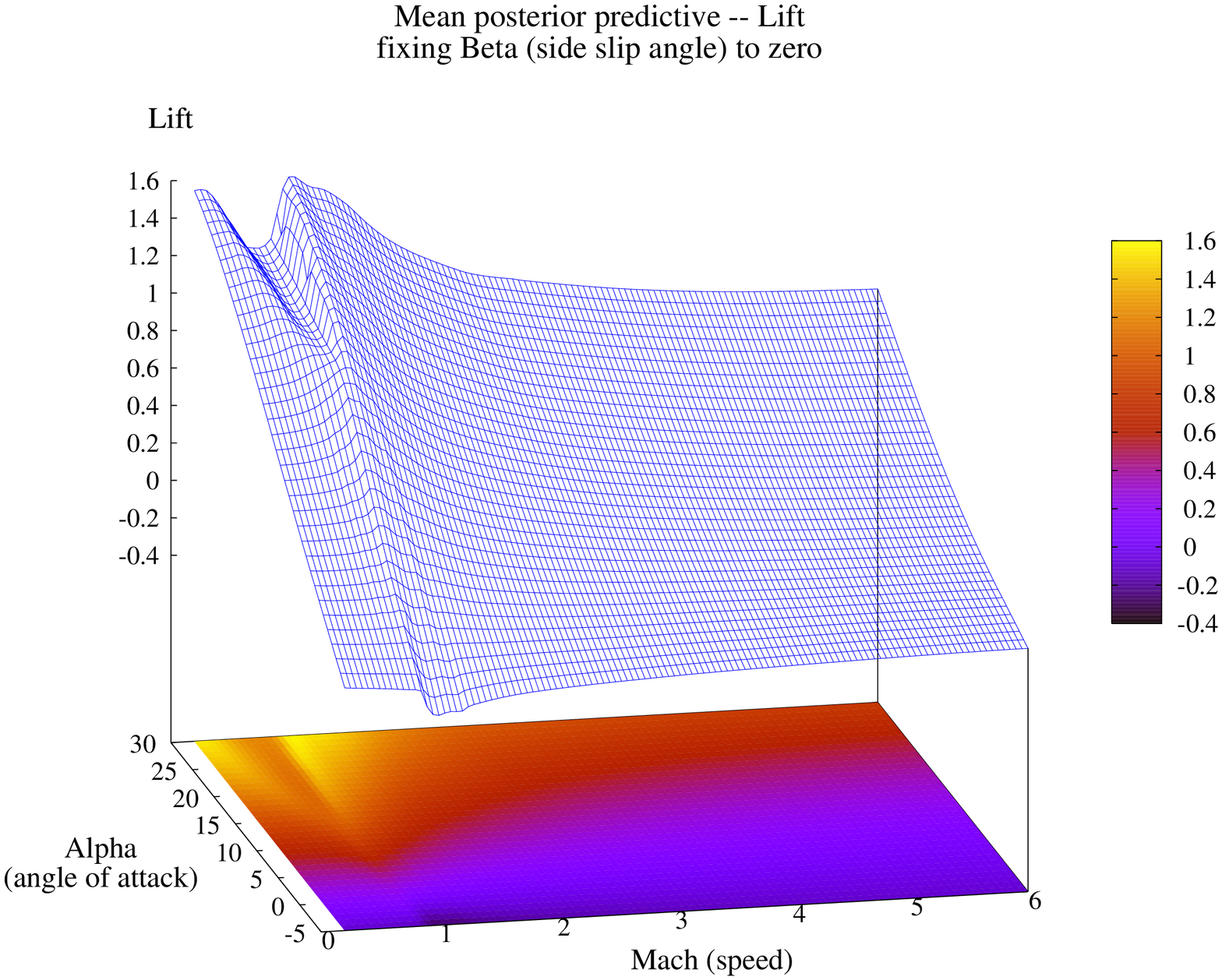}
\includegraphics[scale=0.30,trim=10 0 0 0]{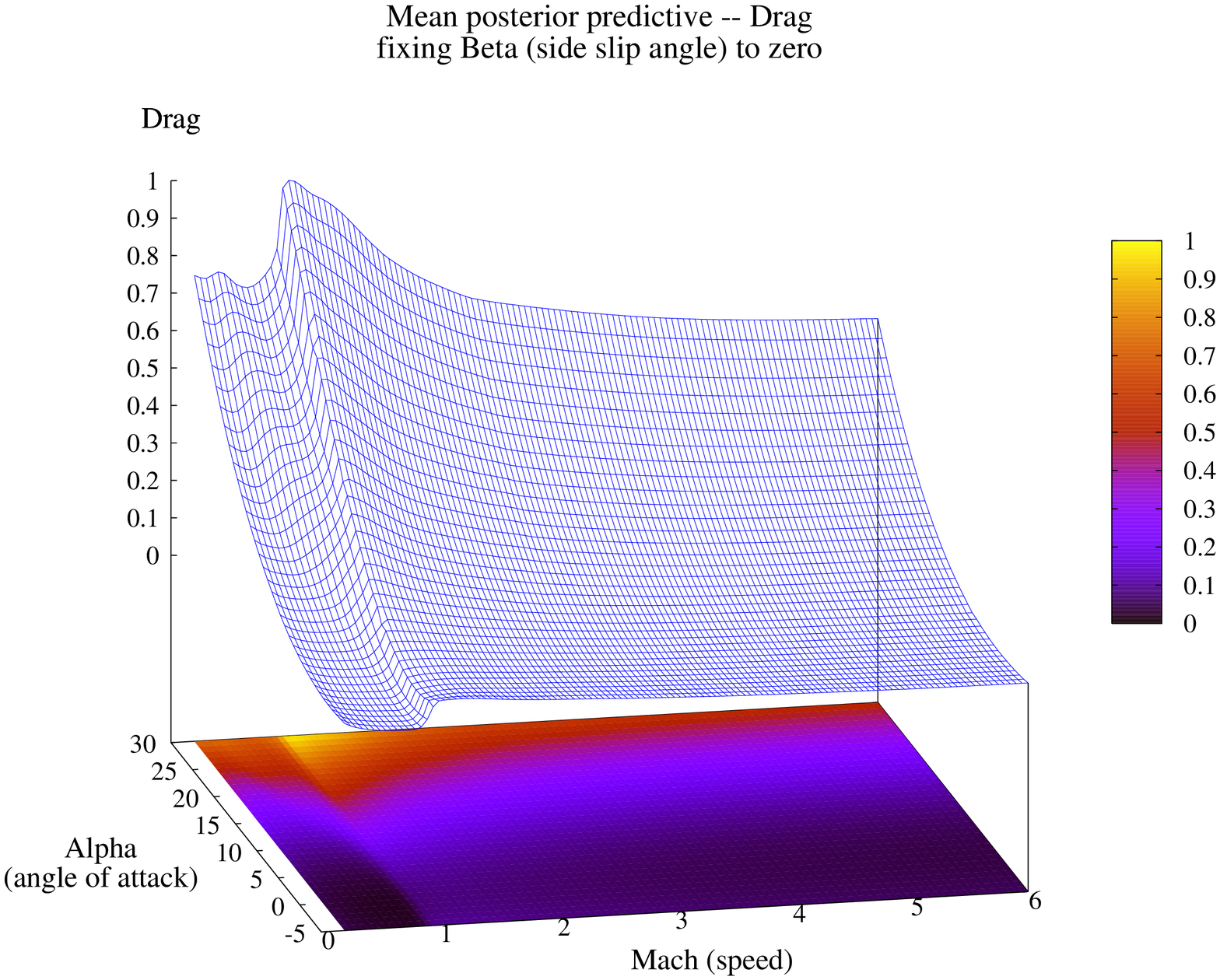}\\
\vspace{-0.4cm}
\includegraphics[scale=0.30,trim=100 0 0 0]{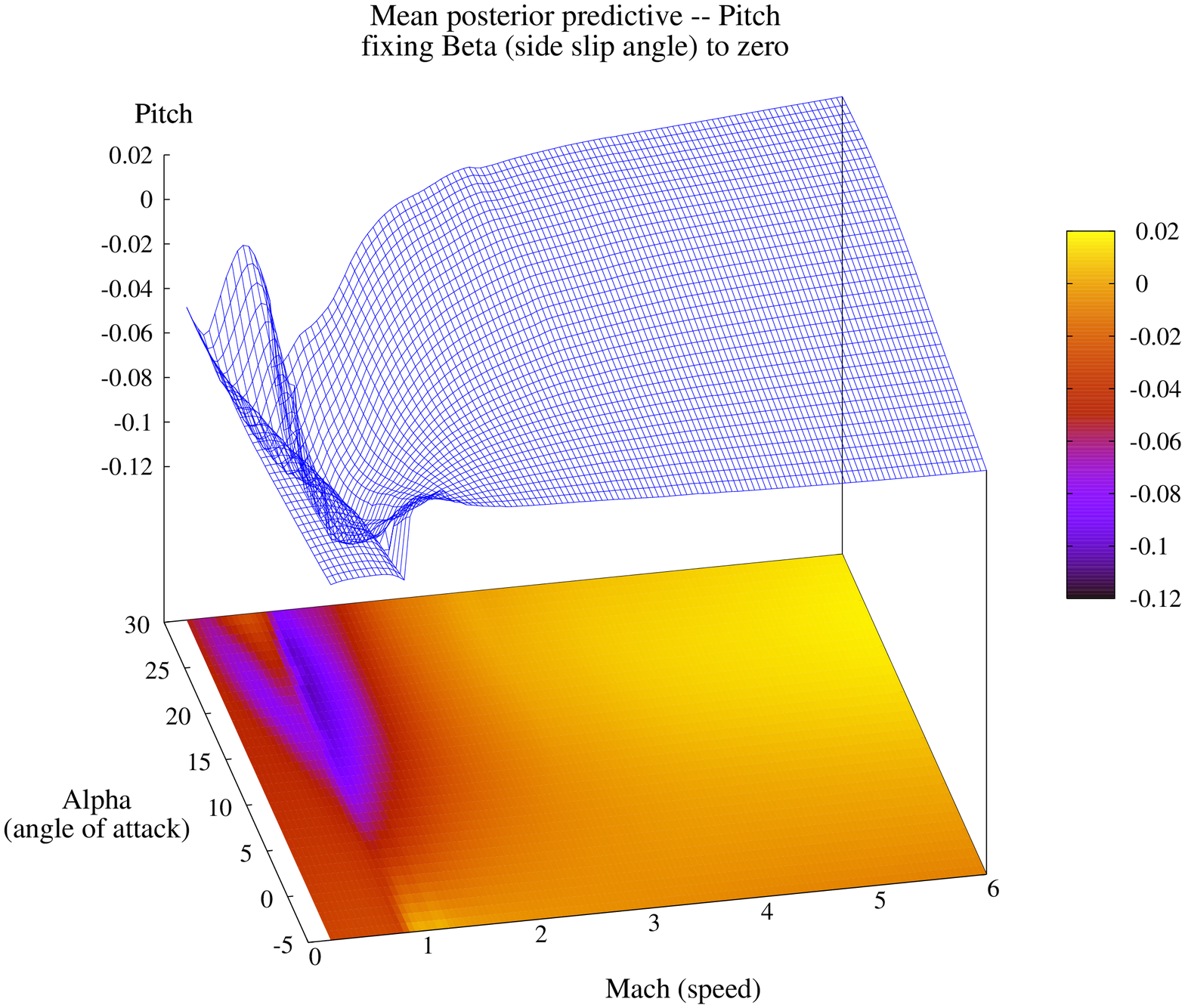}
\includegraphics[scale=0.30,trim=10 0 0 0]{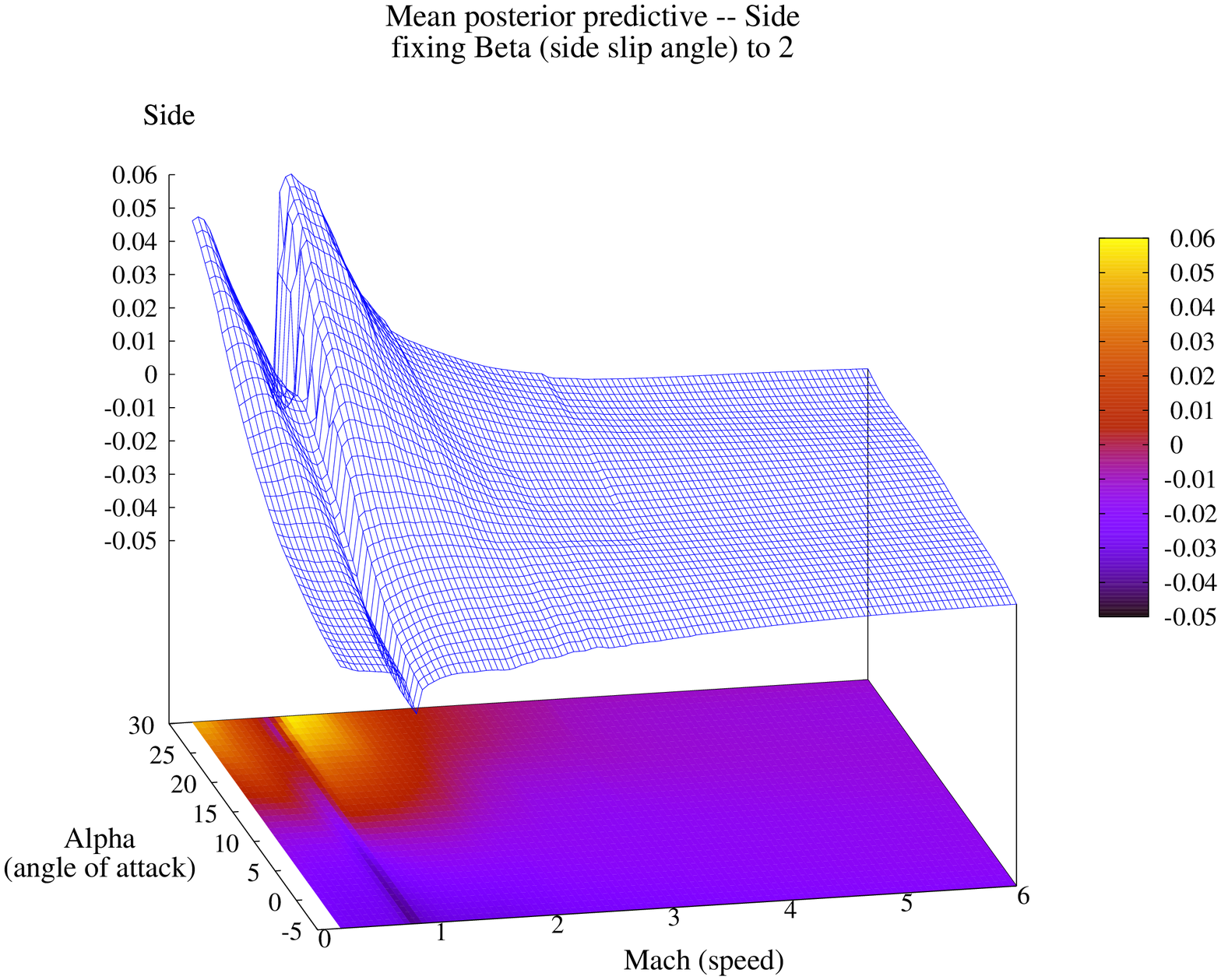}\\
\vspace{-0.4cm}
\includegraphics[scale=0.30,trim=100 0 0 0]{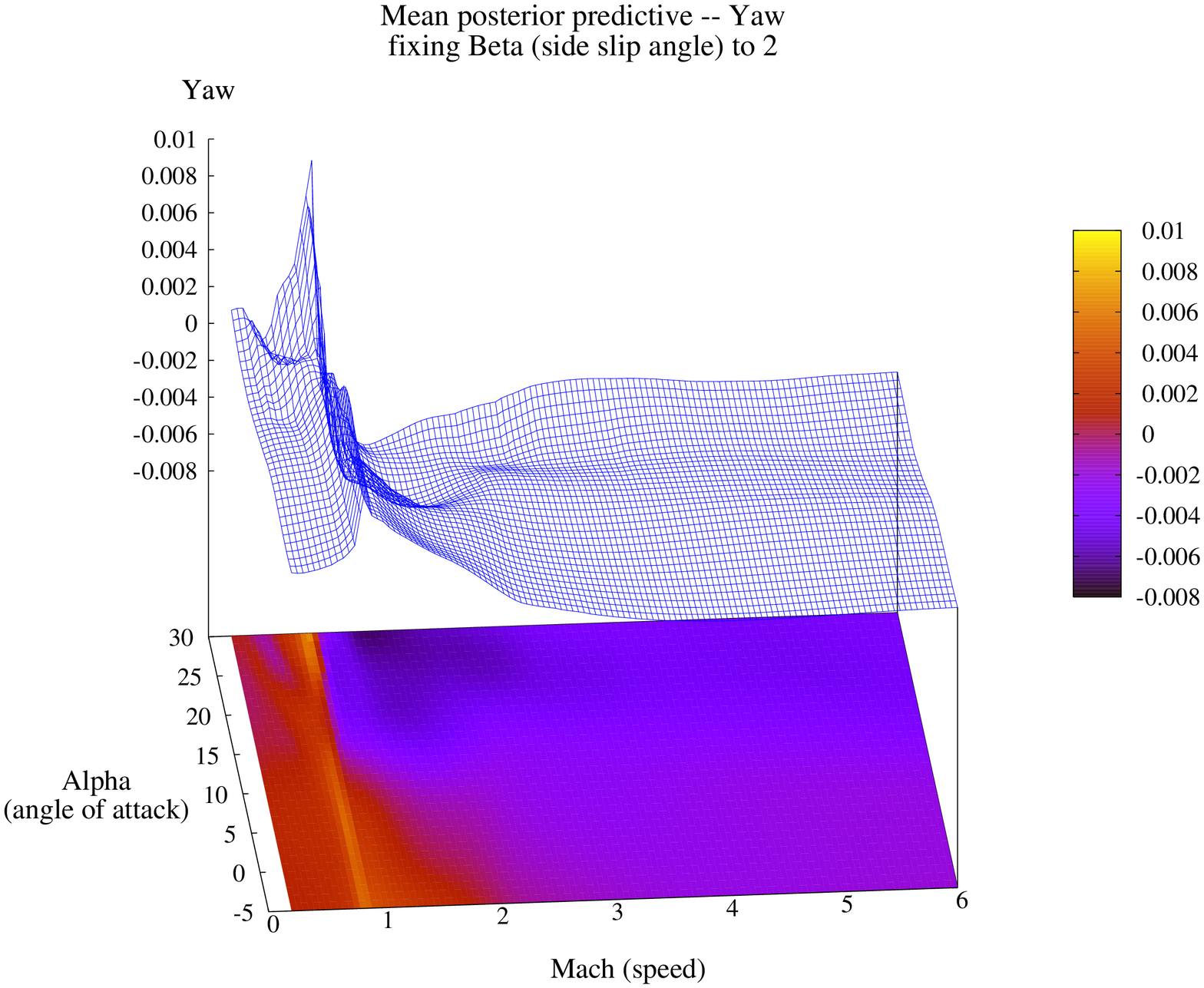}
\includegraphics[scale=0.30,trim=10 0 0 0]{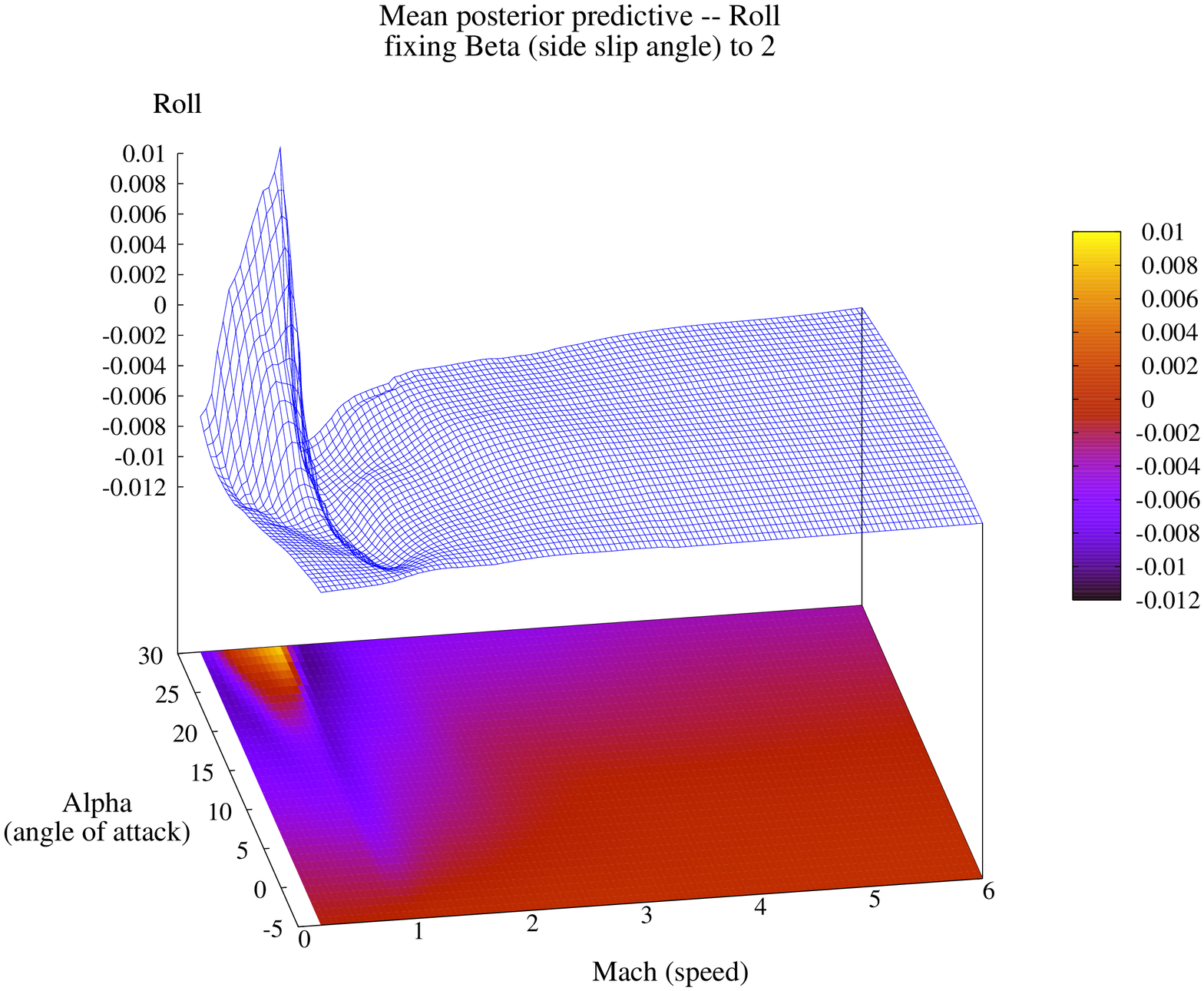}
\vspace{-0.5cm}
\caption[LGBB {\em pitch}, slice Beta = 0, comparison]{LGBB slice of
  mean posterior predictive surface for six responses
  ({\em lift, drag, pitch, side, yaw,
    roll}) plotted as a function of Mach (speed) and Alpha (angle of
  attack) with Beta (side slip angle) fixed at zero for the first
  three responses, and two for the last three.
  \label{f:lgbb}}
\end{center}
\vspace{-0.4cm}
\end{figure}

After samples are gathered, the treed GP model can be used for Monte
Carlo estimation of the posterior predictive distribution.  The {\em
  upper--left} plot in Figure~\ref{f:lgbb} shows a slice of the lift
response, for fixed beta plotted as a function of Mach and alpha.
[The {\em upper--right} panel of Figure~\ref{f:lgbbpoints} shows the
corresponding adaptively sampled configurations and MAP tree
$\hat{\mathcal{T}}$ (for beta~=~0).]  The MAP partition separates out
the near-Mach-one region.  Samples are densely concentrated in this
region---most heavily for large alpha.

Figure~\ref{f:lgbb} shows posterior predictive surfaces for the
remaining five responses as well.  Drag and Pitch are shown for the
beta~=~0 slice.  Other slices look strikingly similar.  Side, yaw, and
roll are shown for the beta~=~2 slice, as beta~=~0 slices for these
responses are essentially zero.  All six responses exhibit similar
characteristics, in that the supersonic cases are tame relative to
their subsonic counterparts, with the most interesting region occurring
near Mach 1, and for large angle of attack (alpha).  The treed
GP model has enabled BAS to key in on this phenomenon and concentrate
sampling there (Figure~\ref{f:lgbbpoints}).  Compared to the
initial experiment, BAS reduced the simulation burden on
the supercomputer by more than 75\%.

\vspace{-0.4cm}
\section{Conclusion}
\label{sec:conclude}
\vspace{-0.3cm}

We showed how the treed GP LLM can be used as a surrogate model in the
sequential design of computer experiments.  A hybrid approach,
combining active learning and classical design methodologies, was
taken in order to develop a flexible system for use in the highly
variable environment of asynchronous agent--based supercomputing.  
Two sampling algorithms were proposed as adaptations to similar
techniques developed for a simpler class of models.  One chooses to
sample configurations with high posterior predictive variance (ALM);
the other uses a criterion based on an average global reduction in
uncertainty (ALC).  These model uncertainty statistics were used to
determine which of a set optimally spaced candidate locations should
go out for simulation next. Optimal candidate designs were determined by
adapting a classic optimal design methodology to Bayesian partition
models.  The result is a highly efficient Bayesian adaptive sampling
(BAS) strategy, representing a significant improvement on the
state-of-the-art of computer experiment methodology at NASA.
ALM, ALC, and treed maximum entropy design are implemented in the {\tt
  tgp} package for {\sf R} available on CRAN.  Code for adaptive
sampling via an asynchronous supercomputer ({\em emcee}) interface is
available upon request.


There are some enhancements which can be made
towards applying the methods herein to a broader array of problems.
Three such closely related problems are of sampling to find extrema
\citep{jones:schonlau:welch:1998}, to find contours
\citep{pritam:2008} generally, or to find boundaries, i.e., contours
with large gradients \citep{banerjee:gelfand:2006}, a.k.a.~Wombling.
Other related problems include that of learning about, or finding
extrema in, computer experiments with multi-fidelity codes of varying
execution costs \citep{huang:allen:notz:miller:2006}, and those which
are paired with a {\em physical} experiment
\citep{reese:wilson:hamada:martz:ryan:2005}.

\vspace{-0.3cm}
\subsection*{Acknowledgments}
\vspace{-0.2cm}

This work was partially supported by research subaward
08008-002-011-000 from the Universities Space Research Association and
NASA, NASA/University Affiliated Research Center grant SC 2003028
NAS2-03144, Sandia National Laboratories grant 496420, and National
Science Foundation grants DMS 0233710 and 0504851.  The authors thank
William Macready for originating the collaboration with NASA and for
his help with the project, Thomas Pulliam and Edward Tejnil for their
help with the NASA data, and the editor(s), associate editor, and two
anonymous referees for their helpful comments and suggestions which
have led to an improved paper.


\renewcommand{\baselinestretch}{1.4}\small\normalsize

\appendix
\input{alc}

\vspace{-0.4cm}
\bibliography{../btgpm/tgp}
\bibliographystyle{../btgpm/jasa}

\end{document}

%% file: alc.tex
\vspace{-0.3cm}
\section{Active Learning -- Cohn (ALC)}
\label{chapt:a:alc}
\vspace{-0.2cm}

Section \ref{sec:a:alcgp} derives ALC for the hierarchical
GP and Section \ref{sec:a:alclm} does the same for the LLM.

\vspace{-0.3cm}
\subsection{For hierarchical Gaussian processes}
\label{sec:a:alcgp}
\vspace{-0.1cm}

The partition inverse equations \citep{barnett:1979} can be used to write
a covariance matrix $\mb{C}_{N+1}$ in terms of $\mb{C}_N$, 
so to obtain an equation for $\mb{C}_{N+1}^{-1}$ in terms of $\mb{C}_N^{-1}$:
\begin{align}
\mb{C}_{N+1} &= \left[ \begin{array}{cc}
\mb{C}_N & \mb{m} \\
\mb{m}^\top & \kappa
\end{array} \right] &
\mb{C}_{N+1}^{-1} &= \left[ \begin{array}{cc}
[\mb{C}_N^{-1} + \mb{g}\mb{g}^\top \mu^{-1} ] & \mb{g} \\
\mb{g}^\top & \mu
\end{array} \right] \label{e:part}
\end{align}
where $\mb{m} = [C(\mb{x}_1, \mb{x}),\dots,C(\mb{x}_N, \mb{x})]$,
$\kappa = C(\mb{x}, \mb{x})$, for an $N+1^{\mbox{\tiny st}}$ point
$\mb{x}$ where $C(\cdot, \cdot)$ is the covariance function, $\mb{g}
= -\mu \mb{C}_N^{-1}\mb{m}$, and $\mu = (\kappa - \mb{m}^\top
\mb{C}_N^{-1}\mb{m})^{-1}$. If $\mb{C}_N^{-1}$ is available, these
partitioned inverse equations allow one to compute
$\mb{C}_{N+1}^{-1}$, without explicitly constructing $\mb{C}_{N+1}$
(in $O(n^2)$ rather than the usual $O(n^3)$).

In the context of ALC sampling from the model in Eq.~(\ref{eq:model}),
the matrix which requires an inverse is $\mb{K}_{N+1} + \mb{F}_{N+1}
\mb{W} \mb{F}_{N+1}^\top$, to calculate the predictive variance
$\hat{\sigma}^2(\mb{x})$.  \vspace{-0.3cm}
\begin{align*}
\mb{K}_{N+1} + \mb{F}_{N+1}^\top \mb{W} \mb{F}_{N+1} &= \left[ \begin{array}{cc}
        \mb{K}_N & \mb{k}_N(\mb{x}) \\
        \mb{k}_N^\top(\mb{x}) & K(\mb{x}, \mb{x})
        \end{array} \right] + \left[ \begin{array}{cc}
        \mb{F}_N \mb{W} \mb{F}_N^\top & \mb{F}_N \mb{W} \mb{f}(\mb{x}) \\
        \mb{f}(\mb{x})^\top \mb{W} \mb{F}_N^\top 
& \mb{f}(\mb{x})^\top \mb{W}\mb{f}(\mb{x}) \end{array} \right] \\
& =  \left[ \begin{array}{cc} 
        \mb{K}_N  + \mb{F}_N \mb{W} \mb{F}_N^\top & \mb{k}_N(\mb{x}) 
                + \mb{F}_N \mb{W} \mb{f}(\mb{x}) \\
        \mb{k}_N^\top(\mb{x}) + \mb{f}(\mb{x})^\top \mb{W} \mb{F}_N^\top & 
                K(\mb{x}, \mb{x}) + \mb{f}(\mb{x})^\top \mb{W}\mb{f}(\mb{x})
        \end{array} \right].
\end{align*}
(*) Using the notation $\mb{C}_N = \mb{K}_N + \mb{F}_N \mb{W} \mb{F}_N^\top$, 
$\mb{q}_N(\mb{x}) = \mb{k}_N(\mb{x}) + \mb{F}_N \mb{W} \mb{f}(\mb{x})$, and 
$\kappa(\mb{x}, \mb{y}) = K(\mb{x}, \mb{y}) 
+ \mb{f}(\mb{x})^\top \mb{W}\mb{f}(\mb{y})$
yields some simplification:
\vspace{-0.3cm}
\[
\mb{C}_{N+1} = \mb{K}_{N+1} + \mb{F}_{N+1} \mb{W} \mb{F}_{N+1}^\top =  
        \left[ \begin{array}{cc} 
        \mb{C}_N & \mb{q}_N(\mb{x}) \\
        \mb{q}_N(\mb{x})^\top & \kappa(\mb{x}, \mb{y}) 
        \end{array} \right].
\]
Applying the partitioned inverse equations (\ref{e:part})
gives
\vspace{-0.3cm}
\begin{equation}
\mb{C}_{N+1}^{-1} = (\mb{K}_{N+1} + \mb{F}_{N+1}^\top \mb{W} \mb{F}_{N+1})^{-1} =
        \left[ \begin{array}{cc}
        [\mb{C}_N^{-1} + \mb{g}\mb{g}^\top\mu^{-1}] & \mb{g} \\
        \mb{g}^\top & \mu 
        \end{array} \right] \label{e:fullinv}
\end{equation}
where $\mb{g} = - \mu \mb{C}_N^{-1} \mb{q}_N(\mb{x})$, and $\mu =
(\kappa(\mb{x}, \mb{x}) -
\mb{q}_N(\mb{x})^\top\mb{C}_N^{-1}\mb{q}_N(\mb{x}))^{-1}$ from (*).
%
We can now calculate the reduction in variance at $\mb{y}$
given that $\mb{x}$ is added into the data:
\vspace{-0.3cm}
\begin{align*}
 && \Delta \hat{\sigma}^2_\mb{y} (\mb{x}) 
        &= \hat{\sigma}^2(\mb{y}) - \hat{\sigma}^2_\mb{x} (\mb{y}), \\
\mbox{where} && 
\hat{\sigma}^2(\mb{y}) &= \sigma^2[\kappa(\mb{y}, \mb{y}) 
        - \mb{q}_N^\top(\mb{y}) \mb{C}_N^{-1} \mb{q}_N^\top(\mb{y})], \\
\mbox{and}  &&
\hat{\sigma}^2_\mb{x}(\mb{y}) &= \sigma^2[\kappa(\mb{y}, \mb{y}) 
        - \mb{q}_{N+1}(\mb{y})^\top \mb{C}_{N+1}^{-1} \mb{q}_{N+1}(\mb{y})].\\
\mbox{Now} && \Delta \hat{\sigma}^2_\mb{y} (\mb{x}) &= 
        \sigma^2[\kappa(\mb{y}, \mb{y}) 
        - \mb{q}_N^\top(\mb{y}) \mb{C}_N^{-1} \mb{q}_N(\mb{y})] 
        -  \sigma^2[\kappa(\mb{y}, \mb{y}) 
        - \mb{q}_{N+1}^\top(\mb{y}) \mb{C}_{N+1}^{-1} \mb{q}_{N+1}(\mb{y})] \\
&&      &= \sigma^2 [ \mb{q}_{N+1}(\mb{y})^\top \mb{C}_{N+1}^{-1} \mb{q}_{N+1}(\mb{y}) 
                - \mb{q}_N^\top(\mb{y}) \mb{C}_N^{-1} \mb{q}_N(\mb{y})].
\end{align*}
Focusing on $\mb{q}_{N+1}^\top(\mb{y}) \mb{C}_{N+1}^{-1} \mb{q}_{N+1}(\mb{y})$, 
first decompose $\mb{q}_{N+1}$:
\vspace{-0.3cm}
\begin{align*}
\mb{q}_{N+1} &= \mb{k}_{N+1}(\mb{y}) + \mb{F}_{N+1} \mb{W} \mb{f}(\mb{y}) \\
        &= \left[ \begin{array}{c} \mb{k}_N(\mb{y}) \\ K(\mb{y}, \mb{x}) 
        \end{array} \right] +
                \left[ \begin{array}{c} \mb{F}_N \\ \mb{f}^\top(\mb{x}) \end{array} \right] 
                \mb{W}\mb{f}(\mb{y})
        = \left[ \begin{array}{c} \mb{k}_N(\mb{y}) + \mb{F}_N\mb{W}\mb{f}(\mb{y})  \\ 
                K(\mb{y}, \mb{x}) + \mb{f}^\top(\mb{x})\mb{W}\mb{f}(\mb{y}) \end{array} \right]
        = \left[ \begin{array}{c} \mb{q}_N(\mb{y})  \\ \kappa(\mb{x}, \mb{y}) \end{array} \right].
\end{align*}
Turning attention back to $\mb{C}_{N+1}^{-1} \mb{q}_{n+1}(\mb{y})$, with
the help of (\ref{e:fullinv}):
\begin{align*}
\mb{C}_{N+1}^{-1} \mb{q}_{N+1}(\mb{y})
&= \left[ \begin{array}{cc}
        \mb{C}_N^{-1} + \mb{g}\mb{g}^\top\mu^{-1} & \mb{g} \\
        \mb{g}^\top & \mu 
        \end{array} \right] \hspace{-0.2cm}
        \left[ \begin{array}{c} \mb{q}_N(\mb{y}) \\ 
        \kappa(\mb{x}, \mb{y}) \end{array} \right]
        = \left[ \begin{array}{c}
                [\mb{C}_N^{-1} + \mb{g}\mb{g}^\top\mu^{-1}] \mb{q}_N(\mb{y}) 
                        + \mb{g} \kappa(\mb{x}, \mb{y}) \\
                \mb{g}^\top \mb{q}_N(\mb{y}) + \mu \kappa(\mb{x}, \mb{y})
        \end{array} \right]
\end{align*}
\vspace{-0.4cm}
\begin{align*}
\mb{q}_{N+1}^\top(\mb{y}) \mb{C}_{N+1}^{-1} \mb{q}_{N+1}(\mb{y}) 
&= \left[ \begin{array}{c} \mb{q}_N(\mb{y})  \\ 
                \kappa(\mb{x}, \mb{y}) \end{array} \right]^\top 
        \left[ \begin{array}{c}
                (\mb{C}_N^{-1} + \mb{g}\mb{g}^\top\mu^{-1}) \mb{q}_N(\mb{y}) 
                        + \mb{g}\kappa(\mb{x}, \mb{y})) \\
                \mb{g}^\top \mb{q}_N(\mb{y}) 
        + \mu \kappa(\mb{x}, \mb{y}) \end{array} \right] \\
        &= \mb{q}_N^\top(\mb{y}) [(\mb{C}_N^{-1} 
        + \mb{g}\mb{g}^\top\mu^{-1}) \mb{q}_N(\mb{y}) 
                + \mb{g} \kappa(\mb{x}, \mb{y})] \\
        &\;\;\;\;\; + \kappa(\mb{x}, \mb{y}) [\mb{g}^\top \mb{q}_N(\mb{y}) 
        + \mu\kappa(\mb{x}, \mb{y})].
\end{align*}
\vspace{-1cm}
\begin{align*}
\mbox{Finally} && \Delta \hat{\sigma}^2_\mb{y} (\mb{x})
        &= \sigma^2 [ \mb{q}_{N+1}(\mb{y})^\top \mb{C}_{N+1}^{-1} \mb{q}_{N+1}(\mb{y}) 
                - \mb{q}_N^\top(\mb{y}) \mb{C}_N^{-1} \mb{q}_N(\mb{y})]. \\
        &&&= \sigma^2 [ 
                \mb{q}_N^\top(\mb{y}) \mb{g} \mb{g}^\top \mu^{-1} \mb{q}_N(\mb{y}) 
                        + 2\kappa(\mb{x}, \mb{y}) \mb{g}^\top \mb{q}_N(\mb{y}) 
                                + \mu \kappa(\mb{x}, \mb{y})^2 ] \\
        &&&= \sigma^2 \mu \left[ 
                \mb{q}_N^\top(\mb{y}) \mb{g} \mu^{-1} - \kappa(\mb{x}, \mb{y})
                \right]^2\\
&& \Delta \hat{\sigma}^2_\mb{y} (\mb{x}) &=
        \frac{\sigma^2 \left[ 
        \mb{q}_N^\top(\mb{y}) \mb{C}_N^{-1} \mb{q}_N(\mb{x}) 
        - \kappa(\mb{x}, \mb{y}) \right]^2}
        {\kappa(\mb{x}, \mb{x}) - \mb{q}_N^\top(\mb{x})\mb{C}_N^{-1}\mb{q}_N(\mb{x})}.
\end{align*}

\vspace{-0.1cm}
\subsection{For hierarchical (limiting) linear models}
\label{sec:a:alclm}
\vspace{-0.1cm}

Under the (limiting) linear model, computing the ALC statistic is more
straightforward.  
\vspace{-0.2cm}
\begin{align*}
\Delta \hat{\sigma}^2_\mb{y} (\mb{x}) 
= \hat{\sigma}^2(\mb{y}) - \hat{\sigma}^2_\mb{x} (\mb{y})
        &= \sigma^2 [ 1 - \mb{f}^\top(\mb{y}) \mb{V}_{\tilde{\beta}_N} \mb{f}(\mb{y}) 
                 - 1 - \mb{f}^\top(\mb{y}) \mb{V}_{\tilde{\beta}_{N+1}} \mb{f}(\mb{y})] \\
        &= \sigma^2 \mb{f}^\top(\mb{y}) [ \mb{V}_{\tilde{\beta}_N}
                - \mb{V}_{\tilde{\beta}_{N+1}} ] \mb{f}(\mb{y}),
\end{align*}
where $\mb{V}_{\tilde{\beta}_{N+1}}$ from \citep{gra:lee:2008}
includes $\mb{x}$, and $\mb{V}_{\tilde{\beta}_N}$ does not.  
Expanding
out $\mb{V}_{\tilde{\beta}_{N+1}}$: \vspace{-0.3cm}
\vspace{-0.3cm}
\begin{align*}
\Delta \hat{\sigma}^2_\mb{y} (\mb{x})
&=\sigma^2 \mb{f}^\top(\mb{y}) \left[ \mb{V}_{\tilde{\beta}_N} - 
        \left(\frac{\mb{W}^{-1}}{\tau^2} 
        + \frac{\mb{F}^\top_{N+1} \mb{F}_{N+1}}{1+g}\right)^{-1}
        \right] \mb{f}^\top(\mb{y}) \\
&=\sigma^2 \mb{f}^\top(\mb{y}) \left[ \mb{V}_{\tilde{\beta}_N} - 
        \left(\frac{\mb{W}^{-1}}{\tau^2} + \frac{1}{1+g}
        \left[ \begin{array}{c} \mb{F}_N \\ \mb{f}^\top(\mb{x}) \end{array} \right]^\top
        \left[ \begin{array}{c} \mb{F}_N \\ \mb{f}^\top(\mb{x}) \end{array} \right]
        \right)^{-1}
        \right] \mb{f}(\mb{y})\\
&=\sigma^2 \mb{f}^\top(\mb{y}) \left[ \mb{V}_{\tilde{\beta}_N} - 
        \left(\frac{\mb{W}^{-1}}{\tau^2} +
        \frac{\mb{F}_N^\top \mb{F}_N}{1+g} 
        + \frac{\mb{f}(\mb{x})\mb{f}^\top(\mb{x})}{1+g} \right)^{-1}
        \right] \mb{f}(\mb{y}) \\
&=\sigma^2 \mb{f}^\top(\mb{y}) \left[ \mb{V}_{\tilde{\beta}_N} - 
        \left(\mb{V}_{\tilde{\beta}_N}^{-1} 
        + \frac{\mb{f}(\mb{x})\mb{f}^\top(\mb{x})}{1+g} \right)^{-1}
        \right] \mb{f}(\mb{y}). \\
\intertext{The Sherman-Morrison-Woodbury formula \citep{berns:2005}, where
$\mb{V} \equiv \mb{f}^\top(\mb{x})(1+g)^{-\frac{1}{2}}$ and $\mb{A} \equiv \
\mb{V}_{\tilde{\beta}_N}^{-1}$  gives \vspace{-0.3cm}}
\Delta \hat{\sigma}^2_\mb{y} (\mb{x}) 
&=\sigma^2 \mb{f}^\top(\mb{y}) \left[ 
        \left(1 + \frac{\mb{f}^\top(\mb{x}) \mb{V}_{\tilde{\beta}_N}\mb{f}(\mb{x})}
        {1+g} \right)^{-1}
        \mb{V}_{\tilde{\beta}_N} \frac{\mb{f}(\mb{x})\mb{f}^\top(\mb{x})}{1+g}
        \mb{V}_{\tilde{\beta}_N} \right]\mb{f}(\mb{y}) \\
\Delta \hat{\sigma}^2_\mb{y} (\mb{x}) &= \frac{ 
        \sigma^2 [\mb{f}^\top(\mb{y}) \mb{V}_{\tilde{\beta}_N} \mb{f}(\mb{x})]^2}
        {1+g + \mb{f}^\top(\mb{x}) \mb{V}_{\tilde{\beta}_N} \mb{f}(\mb{x})}.
\end{align*}